\documentclass[useAMS]{mn2e}

\usepackage{amssymb,amsmath,psfig,times}
\def\gsim{ \lower .75ex \hbox{$\sim$} \llap{\raise .27ex \hbox{$>$}} }
\def\lsim{ \lower .75ex\hbox{$\sim$} \llap{\raise .27ex \hbox{$<$}} }

\def\sc{Schwarzschild}
\def\beq{\begin{equation}}
\def\eeq{\end{equation}}

\def\sc{Schwarzschild}

\title[High redshift Fermi blazars]
{High redshift Fermi blazars}  
\author[G. Ghisellini et al.]
{G. Ghisellini$^1$\thanks{Email: gabriele.ghisellini@brera.inaf.it}, 
G. Tagliaferri$^1$,
L. Foschini$^1$,
G. Ghirlanda$^1$, 
F. Tavecchio$^1$, 
\newauthor{R. Della Ceca$^2$, F. Haardt$^{3,4}$, M. Volonteri$^5$, N. Gehrels$^6$
}
\\
\\
$^1$INAF -- Osservatorio Astronomico di Brera, Via Bianchi 46, I--23807 Merate, Italy\\
$^2$INAF -- Osservatorio Astronomico di Brera, Via Brera 28, I--20100 Milano, Italy\\
$^3$Universit\'a dell'Insubria, Dipartimento di Fisica e Matematica, Via Valleggio 11, 
I--22100 Como, Italy;\\
$^4$ INFN, Sezione di Milano--Bicocca, I--20126 Milano, Italy; \\
$^5$Astronomy Department, University of Michigan, Ann Arbor, MI 48109 \\
$^6$ NASA Goddard Space Flight Center, Greenbelt, USA.
}

\begin{document}  

\maketitle

\begin{abstract}
With the release of the first year {\it Fermi} catalogue,
the number of blazars detected above 100 MeV lying at high redshift
has been largely increased.
There are 28 blazars at $z>2$ in the ``clean" sample.
All of them are Flat Spectrum Radio Quasars (FSRQs).
We study and model their overall spectral energy distribution
in order to find the physical parameters of the jet emitting
region, and for all of them we estimate their black hole masses and
accretion rates. 
We then compare the jet with the accretion disk properties, 
setting these sources in the broader context of all the other
bright $\gamma$--ray or hard X--ray blazars.
We confirm that the jet power correlates with the accretion luminosity.
We find that the high energy emission peak shifts to smaller frequencies
as the observed luminosity increases, according to the blazar sequence, 
making the hard X--ray band 
the most suitable for searching the most luminous and distant blazars.
\end{abstract}

\section{Introduction}

The Large Area Telescope (LAT) onboard the {\it Fermi} satellite
detected, after 11 months of all sky survey, more than 1,400 sources,
presented in Abdo et al. (2010a), with roughly half of them
being BL Lac objects or Flat Spectrum Radio Quasars (FSRQs) 
(Abdo et al. 2010b, hereafter A10) and a few radio--galaxies,
starbursts, galaxies, and Narrow Line Seyfert 1 galaxies.
The corresponding catalog of AGNs at high Galactic latitude 
($|b| > 10^\circ$) is called First LAT AGN Catalog (1LAC).
BL Lacs and FSRQs are approximately present in equal number.

With respect to the previous sample (LAT Bright AGN Sample, hereafter LBAS), 
constructed after 3 months of survey (Abdo et al. 2009; hereafter A09), 
the number of detected blazars is about 7 times larger, 
as a result of the lower limiting sensitivity, obtained with the
longer exposure and the smaller required significance (from 10 $\sigma$
of the first 3 months to the current 4 $\sigma$ level).
Correspondingly, also the number of high redshift blazars
detected in $\gamma$--rays increased: in the LBAS there were 5 blazars at $z>2$
(and none at $z>3$), while in the 1LAC catalogue there are 28 sources at $z>2$
(and 2 at $z>3$).

The increased number of high redshift $\gamma$--ray blazars allows us to 
characterize them in a meaningful way, through their
Spectral Energy Distributions (SEDs) and their modelling.
Indeed, the coverage at other frequencies (besides the {\it Fermi}/LAT band)
includes observations by the {\it Swift} satellite for all sources
both in the optical--UV band (through the Optical--UV Telescope UVOT)
and the soft X--ray band (0.3--10 keV, through the X--Ray Telescope XRT).

It is also interesting to compare the properties of the high redshift 
blazars detected in $\gamma$--rays with the high--$z$ blazars detected in
hard X--rays by the Burst Alert Telescope (BAT) instrument 
onboard the {\it Swift} satellite.
All blazars at $z>2$ are FSRQs, so, up to now, high redshift
``blazars" coincide with high redshift FSRQs, since no BL Lac objects
with a measured redshifts $z>2$ has been detected so far.
There are 10 FSRQs at $z>2$ and 5 at $z>3$ in the 3--year BAT all sky survey 
presented by Ajello et al. (2009), that have been studied in Ghisellini et al. (2010a, hereafter G10).
The BAT and the LAT samples of high redshift blazars are rather well defined,
since the sky coverage is quasi--uniform (excluding the Galactic plane)
and we can consider these samples as flux limited.

The main aims of the present paper are then to characterize the properties of 
blazars detected at high energies
at redshift greater than 2 and to see if we can understand the differences (if any)
between the blazars detected in the two bands ($\gamma$--rays and hard X--rays).
In G10, in fact, we suggested that the best way to select the most powerful
blazars at large redshifts is through a survey in the hard X--ray band,
rather than in the $\gamma$--ray one, but this was based on small numbers.
None of the 10 BAT blazars at $z>2$ is present in the LBAS catalogue, 
and only 4 of them are in the 1LAC sample, i.e. have been detected after 
11 months of survey by the {\it Fermi}/LAT instrument.
In G10 we explained this through a change of the average SED when
the bolometric luminosity changes: by increasing it, the high energy
hump of the SED peaks at smaller frequencies, in agreement with the
blazar sequence as put forward by Fossati et al. (1998) and Donato et al. (2001),
and interpreted in Ghisellini et al. (1998).

We anticipate that our earlier suggestion remains valid, with important
implications on the planned future hard X--ray survey missions, such as {\it EXIST}.

In this paper we use a cosmology with $h=\Omega_\Lambda=0.7$ and $\Omega_{\rm M}=0.3$,
and use the notation $Q=10^X Q_x$ in cgs units (except for the black hole masses,
measured in solar mass units).

\section{The high redshift sample}

We consider all blazars detected during the first year all--sky
survey of {\it Fermi} and classified as ``clean" in the catalogue of A10. 
These are all the blazars with $|b|>10^\circ$, detected at more 
than the 4$\sigma$ level whose identification is secure and unique.
In total the 1LAC clean sample contains 599 sources (A10), of which 248 
are FSRQs, all with a measured redshift, and 275 BL Lacs (116 with the redshift
measured).
Among these, we selected the 27 blazars at $z>2$
as listed and classified by A10, plus an additional
source, SWIFT J1656.3--3302 ($z=2.4$), that Ghirlanda et al. (2010) recently
classified as FSRQ among the unidentified 1LAC sources.

Five of these were already present in the LBAS list, i.e.
the blazars detected at more than the 10$\sigma$ level during the
first 3 months of {\it Fermi} survey (A09).
Four additional sources are present in the 3--years survey
of {\it Swift}/BAT (A09) and they too have been studied in G10.

Table \ref{sample} lists all sources: the top 19 blazars are
studied in this paper, while the bottom 9 are the sources
already present either in the BAT or LBAS samples.
In this paper we present the spectral energy distributions (SED)
and the modelling for the ``new" ones, i.e. blazars not present
in our previous study (G10).

\begin{table} 
\centering
\begin{tabular}{lllll}
\hline
\hline
Name        &Alias   &$z$    &$\log L_\gamma$ &Ref \\
\hline 
0106+01    &4C+01.02 &2.107  &48.7  &0   \\
0157--4614 &PMN      &2.287  &47.9  &0   \\
0242+23    &B2       &2.243  &48.0  &0  \\ 
0322+222   &TXS      &2.066  &48.0  &0  \\ 
0420+022   &PKS      &2.277  &47.9  &0  \\    
0451--28   &PKS      &2.56   &48.7  &0 \\ 
0458--02   &PKS      &2.291  &48.1  &0 \\ 
0601--70   &PKS      &2.409  &48.3  &0  \\ 
0625--5438 &PMN      &2.051  &48.2  &0 \\    
0907+230   &TXS      &2.661  &48.3  &0 \\ 
0908+416   &TXS      &2.563  &47.7  &0 \\ 
1149--084  &PKS      &2.367  &47.7  &0  \\ 
1343+451   &TXS      &2.534  &48.4  &0  \\ 
1344--1723 &PMN      &2.49   &48.5  &0  \\    
1537+2754  &[WB92]   &2.19   &47.6  &0 \\ 
1656.3--3302 &Swift  &2.4    &48.1  &0 \\
1959--4246 &PMN      &2.174  &48.0  &0 \\ 
2118+188   &TXS      &2.18   &48.1  &0 \\    
2135--5006 &PMN      &2.181  &48.1  &0 \\ 
\hline
0227--369  &PKS      &2.115  &48.1  &G09 \\ 
0347--211  &PKS      &2.944  &49.1  &G09 \\ 
0528+134   &PKS      &2.07   &48.8  &G09 \\    
0537-286   &PKS      &3.104  &48.4  &G10 \\ 
0743+259   &TXS      &2.979  &48.6  &G10 \\ 
0805+6144  &CGRaBS   &3.033  &48.4  &G10 \\
0836+710   &4C+71.07 &2.218  &48.5  &G10 \\
0917+449   &TXS      &2.19   &48.4  &G09 \\
1329-049   &PKS      &2.15   &48.5  &G09 \\
\hline
\hline 
\end{tabular}
\vskip 0.4 true cm
\caption{
List of blazars at $z>2$.
The upper part of the table reports the blazars studied in this paper 
(denoted by ``0" in the last column); while the bottom part
reports blazars studied previously:
in Ghisellini et al. (2010; G10) (blazars with $z>2$ detected by {\it Swift}/BAT); 
and in Ghisellini, Tavecchio \& Ghirlanda (2009; G09) (blazars with $z>2$ in LBAS with  
$L_\gamma>10^{48}$erg s$^{-1}$).
}
\label{sample}
\end{table}

\begin{table*}
 \centering
 \begin{minipage}{140mm}
  \begin{tabular}{lccccccc}
\hline
Name & Obs Date & Exp & $N_{\rm H}$ & $\Gamma$ & $F_{0.2-10\rm keV}^{\rm obs}$ & ${\chi^2}$/Cash & d.o.f. \\ 
{}   & DD/MM/YYYY & [ks] & [$10^{20}$~cm$^{-2}$] & {} & [$10^{-13}$~erg~cm$^{-2}$~s$^{-1}$] & {} & {}\\
\hline
0106+01      & 2007--2008\footnote{sum of observations of: 02/07/2007, 10/01/2008, 16/02/2008, 16/08/2009.} & 14.8 & 2.32 & $1.6\pm 0.1$ & $12.8\pm 0.7$ & 1.57/--- & 10\\
0157--4614   & 02/05/2010 & 4.8 & 1.92 & $1.0\pm 0.9$ & $2.1\pm 0.6$ & ---/1.6 & 4\\
0242+23      & 11/03/2010 & 1.2 & 9.46 & $1.5\pm 0.6$ & $7.6\pm 1.9$ & ---/3.7 & 6\\
0322+222     & 25/03/2007 & 2.8 & 8.90 & $1.1\pm 0.2$ & $33\pm 3$ & ---/48.99 & 55\\
0420+022     & 27/03/2010 & 4.2 & 10.7 & $1.8\pm 0.7$ & $3.0\pm 0.6$ & ---/6.7 & 9\\
0451--28     & 27/10/2009 & 6.6 & 2.00 & $1.6\pm 0.1$ & $39\pm 2$ & 0.91/--- & 22\\
0458--02     & 2007--2009\footnote{sum of observations of: 22/03/2007, 10/04/2007, 08/08/2007, 10/08/2007, 13/01/2008, 20/04/2008, 22/04/2008, 26/10/2008, 22/04/2009.} & 31.1 & 5.97 & $1.52\pm 0.08$ & $14.3\pm 0.5$ & 1.19/--- & 24\\
0601--70     & 2008--2009\footnote{sum of observations of: 26/12/2008, 08/01/2009.} & 9.0 & 11.1 & $2.0\pm 0.3$ & $6.3\pm 0.6$ & 1.16/--- & 3 \\
0625--5438   & 03/04/2010 & 5.2 & 7.60 & $1.2\pm 0.3$ & $9.5\pm 1.1$ & ---/37.8 & 33\\
0907+230     & 30/12/2009 & 7.8 & 4.83 & $1.5\pm 0.5$ & $2.4\pm 0.4$ & ---/5.7 & 13\\
0908+416     & 2010\footnote{sum of observations of: 21/02/2010, 25/02/2010.} & 4.6 & 1.64 & $2.1\pm 0.4$ & $2.8\pm 0.5$ & ---/59.5 & 12\\
1149--084    & 10/11/2009 & 1.0 & 4.75 & $1.6\pm 1.0$ & $5.1\pm 1.5$ & ---/0.85 & 2\\
1343+451     & 2009\footnote{sum of observations of: 06/03/2009, 01/10/2009, 04/10/2009.} & 11.5 & 1.91 & $1.2\pm 0.3$ & $5.1\pm 0.5$ & ---/36.26 & 41\\
1344--1723\footnote{Flux derived by using WebPIMMS with a rate of $(5\pm 2)\times 10^{-3}$~c/s and fixed parameters.}    & 29/12/2009 & 1.0 & 8.70 & $2.0$ & $1.9\pm 0.8$ & ---/--- & ---\\
1539+2744    & 17/03/2010 & 7.1 & 2.81 & $1.4\pm 0.5$ & $4.0\pm 0.6$ & ---/24.6 &19\\
1656--3302   & 2006\footnote{sum of observations of: 09/06/2006, 13/06/2006.} & 9.0 & 22.2 & $1.2\pm 0.1$ & $62\pm 2$ & 0.73/--- & 29\\
1959--4246   & 04/04/2010 & 5.0 & 4.82 & $1.5\pm 0.3$ & $8.2\pm 0.9$ & ---/32.5 & 30\\
2118+188     & 2009\footnote{sum of observations of: 08/01/2009, 13/01/2009.} & 22.8 & 5.36 & $1.9\pm 0.5$ & $2.2\pm 0.2$ & 0.60/--- & 2 \\
2135--5006   & 22/04/2010 & 4.2 & 2.04 & $1.0\pm 0.6$ & $5.3\pm 1.0$ & ---/12.4 & 11\\
\hline
\end{tabular}
\caption{
Summary of XRT observations.
The observation date column indicates the date of a single snapshot or the years during
which multiple snapshots were performed. 
The corresponding note reports the complete set of observations integrated. 
The column ``Exp" indicates the
effective exposure in ks, while $N_{\rm H}$ is the Galactic absorption
column in units of [$10^{20}$ cm$^{-2}$] from Kalberla et al. (2005).
$\Gamma$ is the photon index of the power law model [$F(E)\propto E^{-\Gamma}$], 
$F_{0.2-10\rm keV}^{\rm obs}$ is the observed (absorbed) flux.
The two last columns indicate the
results of the statistical analysis: the last column contains the degrees
of freedom, while the last but one column displays the reduced $\chi^2$ or
the value of the likelihood (Cash 1979), in the case there were no
sufficient counts to apply the $\chi^2$ test.
}
\end{minipage}
\label{xrt}
\end{table*}

\begin{table*}
\centering
\begin{tabular}{llllllll}
\hline
\hline
Source     &$A_V$  &$v$              &$b$             &$u$            &$uvw1$         &$uvm2$           &$uvw2$\\
\hline
0106+01    &0.08   &$17.98\pm 0.12$ &$18.39\pm 0.07$ &$17.63\pm 0.06$ &$18.77\pm0.1$   &$19.75\pm 0.16$  &$20.62\pm 0.22$\\
0157--4614 &0.072  &$>20.37$        &$>21.26$        &$20.52\pm 0.25$ &...             &...              &... \\
0242+23    &0.713  &...             &...             &...             &...             &$>20.66$         &...  \\
0322+222   &0.722  &$>18.61$        &$>19.57$        &$>19.29$        &$>19.79$        &$>20.18$         &$>20.53$ \\
0420+022   &0.719  &$>19.08$        &$19.93\pm 0.34$ &$19.30\pm 0.26$ &$19.89\pm 0.31$ &$20.13\pm 0.33$  &$20.00\pm 0.21$ \\
0451--28   &0.105  &...             &...             &...             &$>20.43$        &...              &...       \\
0458--02   &0.251  &$19.08\pm 0.22$ &$19.51\pm 0.14$ &$19.99\pm 0.26$ &...             &...              &...       \\
0601--70   &0.249  &$19.22\pm 0.22$ &$19.89\pm 0.15$ &$20.10\pm 0.25$ &$20.55\pm 0.28$ &...              &...       \\
0625--5438 &0.472  &$19.31\pm 0.21$ &$19.66\pm 0.18$ &$18.56\pm 0.11$ &$18.96\pm 0.11$ &$19.89\pm 0.19$  &$21.10\pm 0.33$ \\
0907+230   &0.163  &...             &...             &...             &$>20.56$        &...              &...             \\
0908+416   &0.056  &$>19.52$        &$>20.46$        &$>20.15$        &$>20.52$        &$>20.57$         &$>21.23$        \\
1149--084  &0.227  &...             &$>19.16$        &$18.84\pm 0.24$ &$19.98\pm 0.36$ &...              &...             \\
1343+451   &0.078  &$>19.67$        &$>20.60$        &$>20.26$        &$>20.68$        &$>20.91$         &$>21.44$        \\
1344--1723 &0.369  &...	           &...             &...             &...             &$>20.59$         &...             \\
1537+2744  &0.094  &$19.16\pm 0.2$  &$19.63\pm 0.11$ &$19.23\pm 0.11$ &$19.86\pm 0.14$ &$20.11\pm 0.18$  &$19.88\pm 0.10$ \\
1656--3302 &2.09   &...             &...             &$>20.1$         &$>20.47$        &...              &...  \\
1959--4246 &0.259  &$18.52\pm 0.07$ &$19.18\pm 0.06$ &$19.07\pm 0.07$  &...             &...             &...              \\
2118+188   &0.393  &$>20.08$        &$20.27\pm 0.2$  &$19.75\pm 0.18$  &$20.52\pm 0.23$ &$>21.29$        &$>21.77$         \\
2135--5006 &0.078  &$>19.82$        &$>20.17$        &$>20.05$         &...             &...             &...              \\
\hline
\hline
\end{tabular}
\caption{Summary of \emph{Swift}/UVOT observed magnitudes. Lower limits are at $3\sigma$ level.}
\label{uvot}
\end{table*} 

\section{Swift observations and analysis}

For all blazars studied in this paper there are {\it Swift} observations.
Even when they were performed during the 11 months of the 1LAC survey,
they correspond to a ``snapshot" of the optical--X--ray state of the source,
while the $\gamma$--ray data are an average over the 11 months.
Given the very rapid blazar variability, the SEDs constructed in this way
should be considered, in all cases, not simultaneous (but the {\it Swift} UVOT
and XRT data are indeed simultaneous).


The data were screened, cleaned and analysed with the software package
HEASOFT v. 6.8, with the calibration database updated to 30 December 2009.
The XRT data were processed with the standard procedures ({\texttt{XRTPIPELINE v.0.12.4}). 
All sources were observed in photon counting (PC) mode and grade 0--12 
(single to quadruple pixel) were selected. 
The channels with energies below 0.2 keV and above 10 keV were excluded from the fit 
and the spectra were rebinned in energy so to have at least 20--30 counts per bin
in order to apply the $\chi^2$ test. 
When there are no sufficient counts,
then we applied the likelihood statistic in the form reported by Cash (1979). 
Each spectrum was analysed through XSPEC v. 12.5.1n
with an absorbed power law model with a fixed Galactic column density as
measured by Kalberla et al. (2005). 
The computed errors represent the 90\% confidence interval on the spectral parameters. 
Tab. \ref{xrt} reports the log of the observations and the best fit results of 
the X--ray data with a simple power law model. 
The X--ray spectra displayed in the SED have been properly rebinned to ensure the best 
visualization.

UVOT (Roming et al. 2005) source counts were extracted from 
a circular region $5''-$sized centred on the source position, 
while the background was extracted from 
a larger circular nearby source--free region.
Data were integrated with the \texttt{uvotimsum} task and then 
analysed by using the  \texttt{uvotsource} task.
The observed magnitudes have been dereddened according to the formulae 
by Cardelli et al. (1989) and converted into fluxes by using standard 
formulae and zero points from Poole et al. (2008).
Tab. \ref{uvot} lists the observed magnitudes in the 6 filters of UVOT.

\begin{figure}
\vskip -0.6cm \hskip -0.4 cm
\psfig{figure=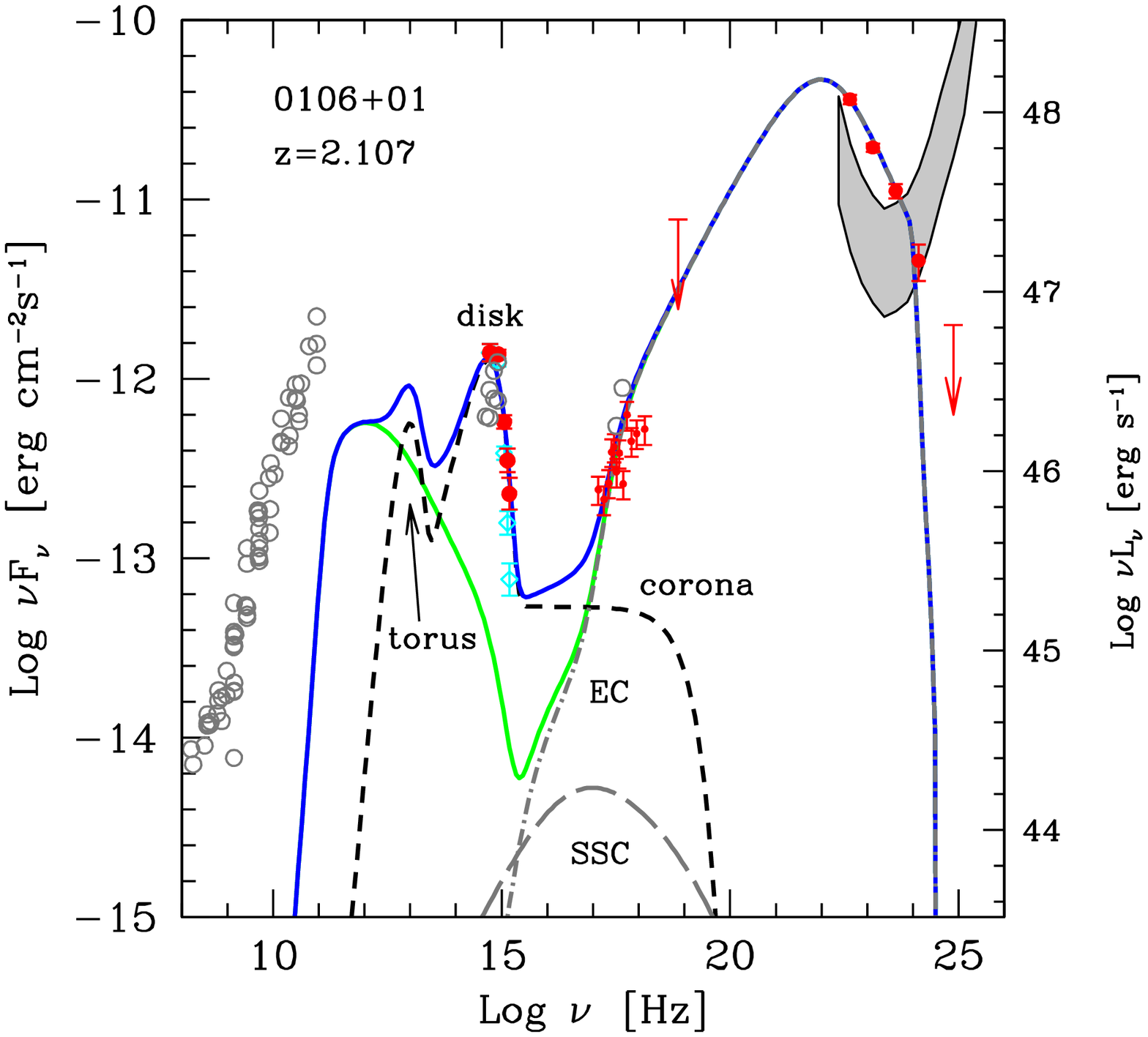,width=9cm,height=9cm}
\vskip -0.8 cm
\caption{
SEDs of 0106+01 (=4C+01.02)
together with the fitting models,
with parameters listed in Tab. \ref{para}.
De--absorbed UVOT, XRT and BAT data are indicated
by darker symbols (red in the electronic version),
while archival data (from NED) are in light grey.
Diamonds (and lower arrows, cyan in the electronic version) 
indicate UVOT data not de--absorbed by
intervening Lyman--$\alpha$ clouds.
The short--dashed line is the emission from the IR torus, 
the accretion disk and its X--ray corona. 
The solid thin (green) line is the non--thermal emission
(sum of synchrotron, SSC and EC).
The long dashed and the dot--dashed grey lines are 
the synchrotron self--Compton 
(SSC) and the External Compton (EC) components, respectively. 
The thick solid (blue) line is the sum of all components.
The grey stripe in the $\gamma$--ray band corresponds to the
{\it Fermi}/LAT
sensitivity of the first 3 months (10$\sigma$, upper boundary)
and for 11 months (4$\sigma$, lower boundary).
}
\label{f0}
\end{figure}

\begin{figure}
\vskip -0.6cm \hskip -0.4 cm
\psfig{figure=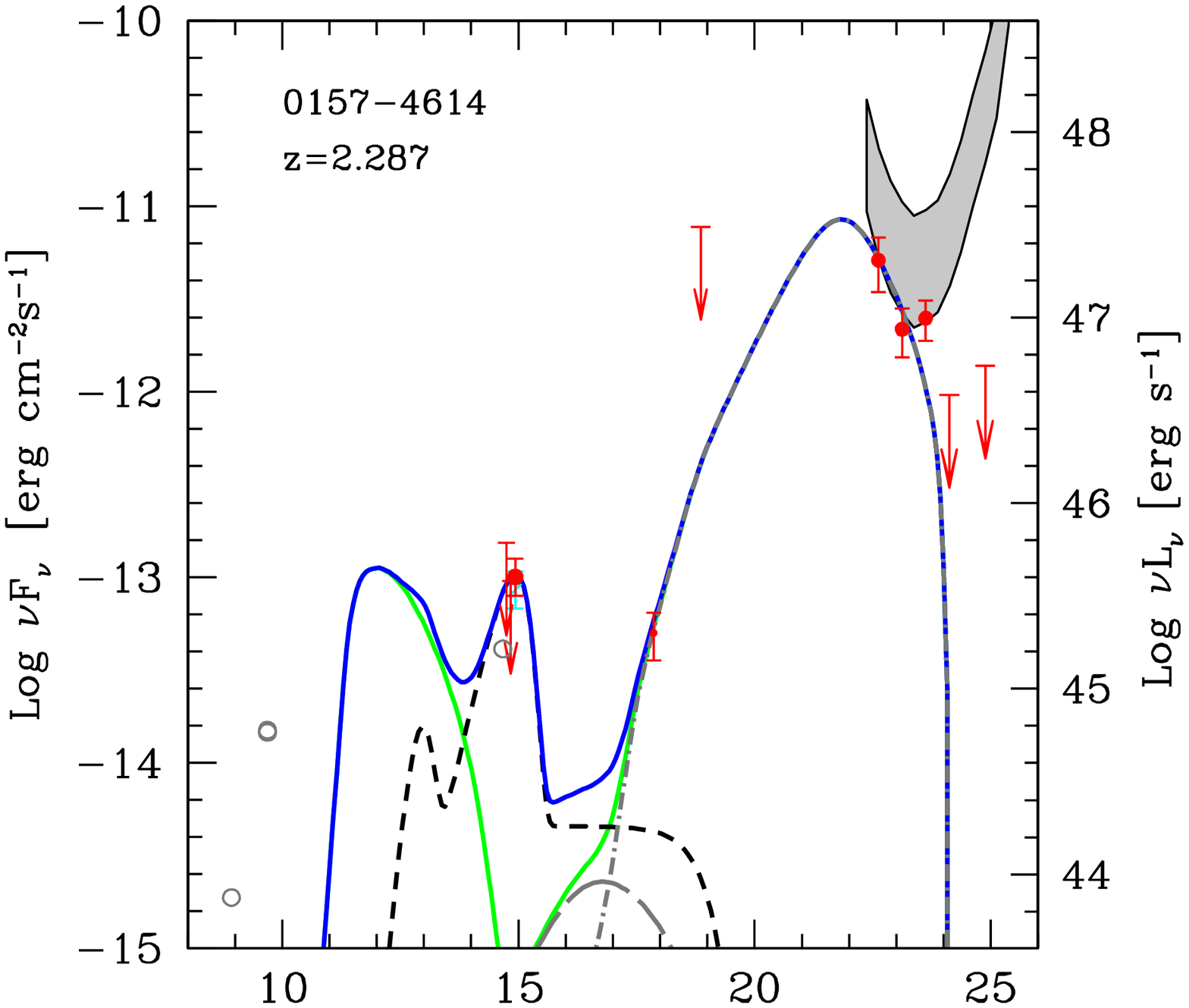,width=9cm,height=6.9cm}
\vskip -1.3 cm \hskip -0.4 cm
\psfig{figure=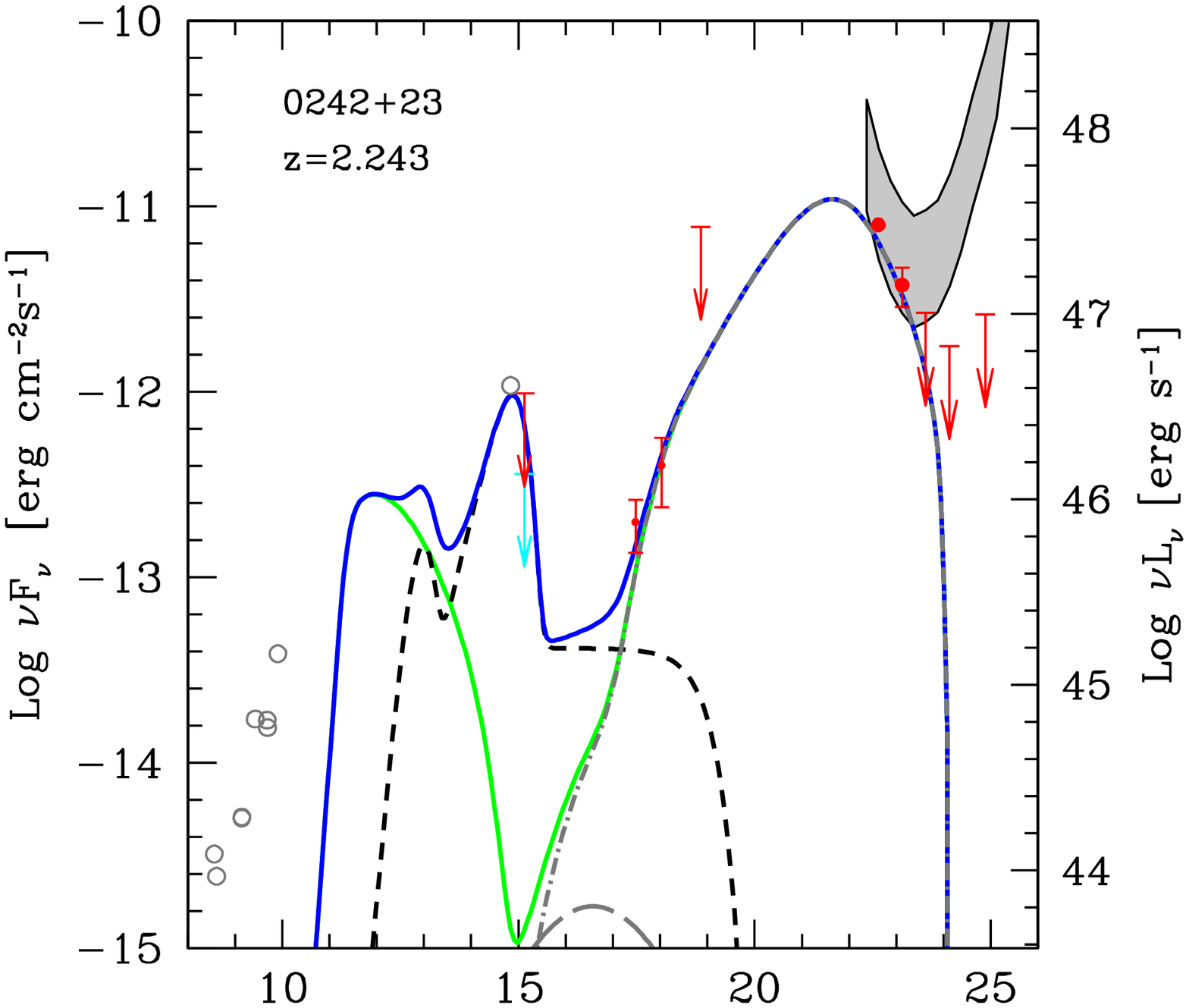,width=9cm,height=6.9cm}
\vskip -1.3 cm \hskip -0.4 cm
\psfig{figure=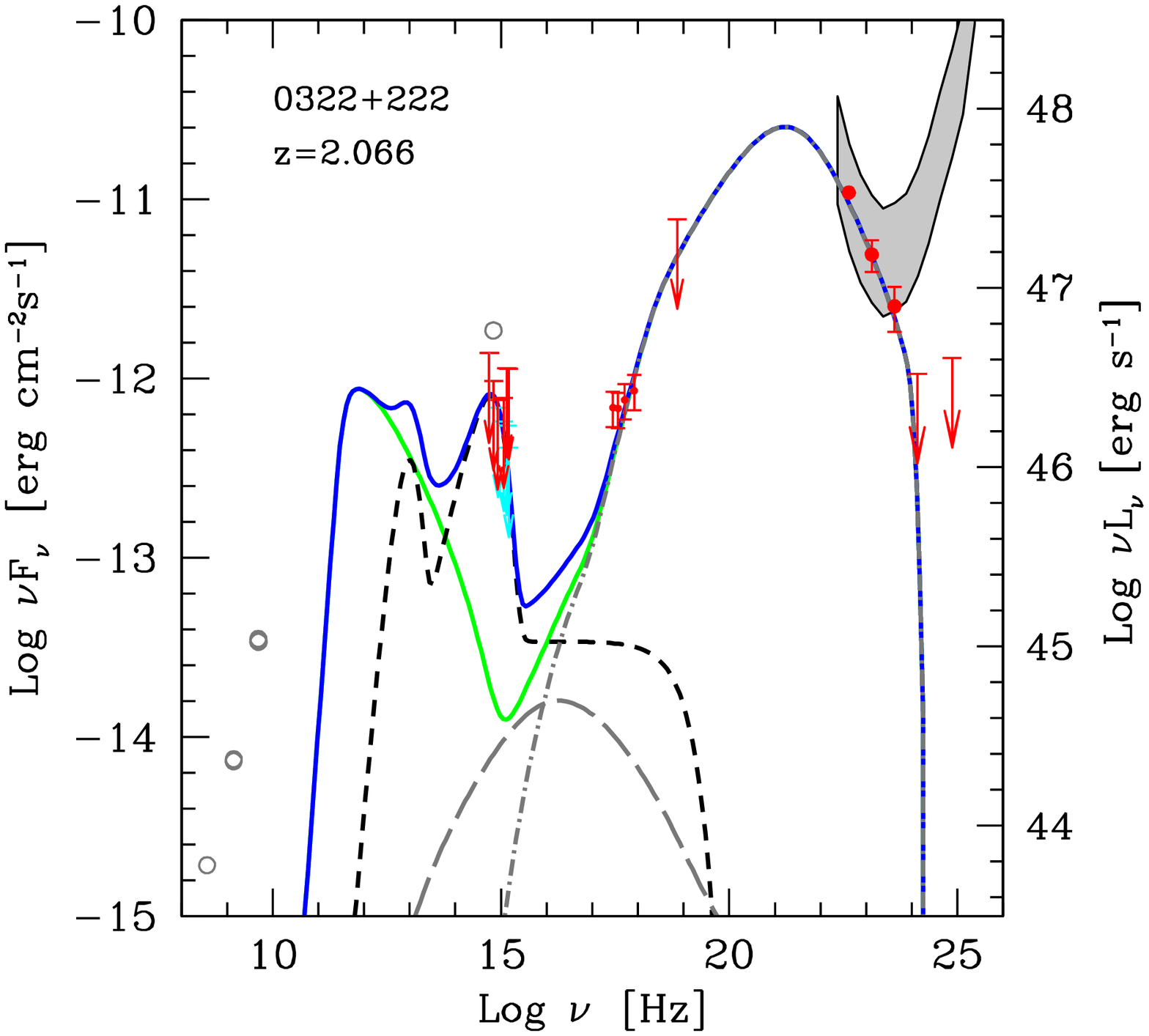,width=9cm,height=6.9cm}
\vskip -0.8 cm
\caption{SED of PMN 0157--4614, B2 0242+23 and TXS 0322+222.
Symbols and lines as in Fig. \ref{f1}.
}
\label{f1}
\end{figure}

\begin{figure}
\vskip -0.6 cm \hskip -0.4 cm
\psfig{figure=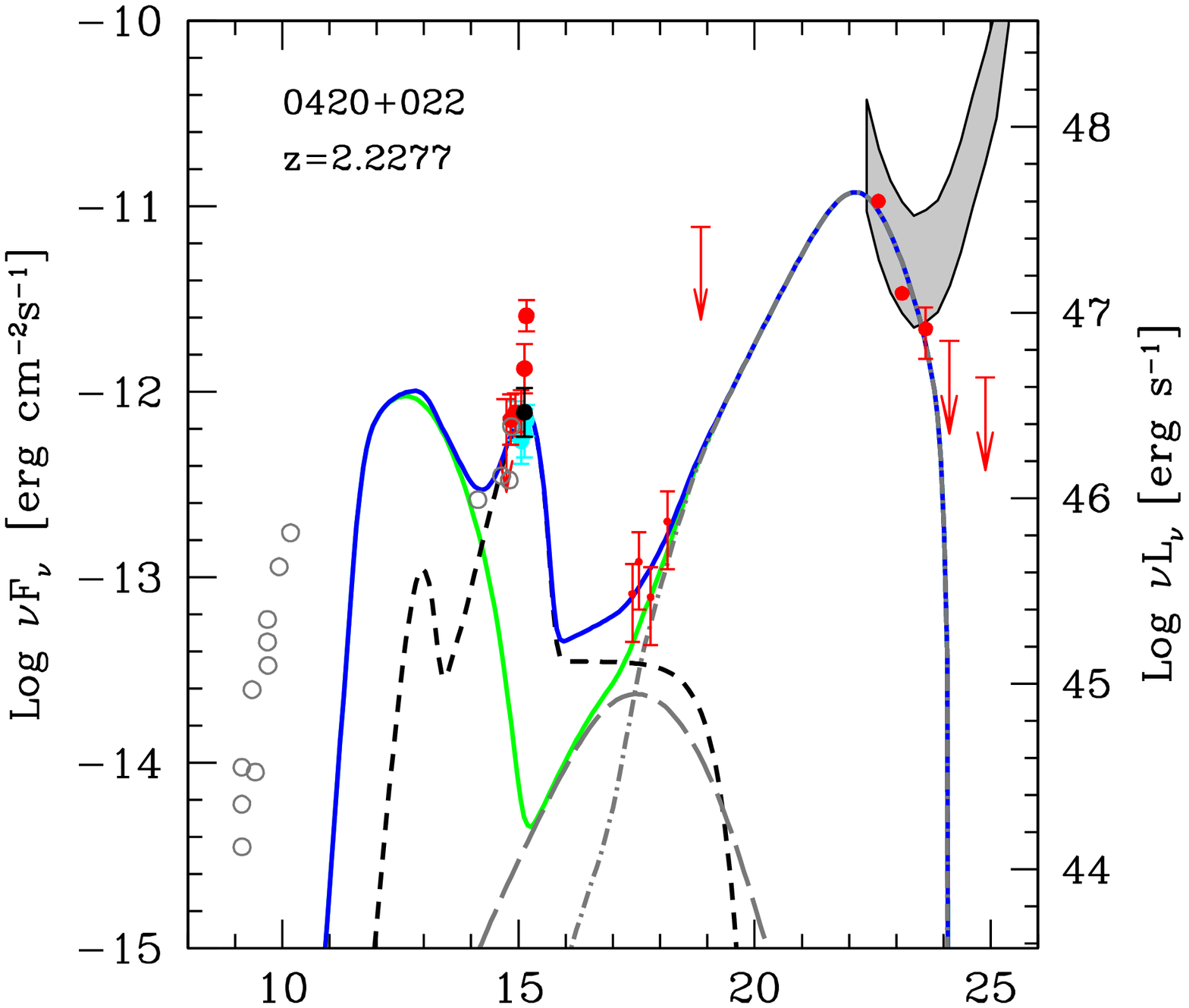,width=9cm,height=7cm}
\vskip -1.3 cm \hskip -0.4 cm
\psfig{figure=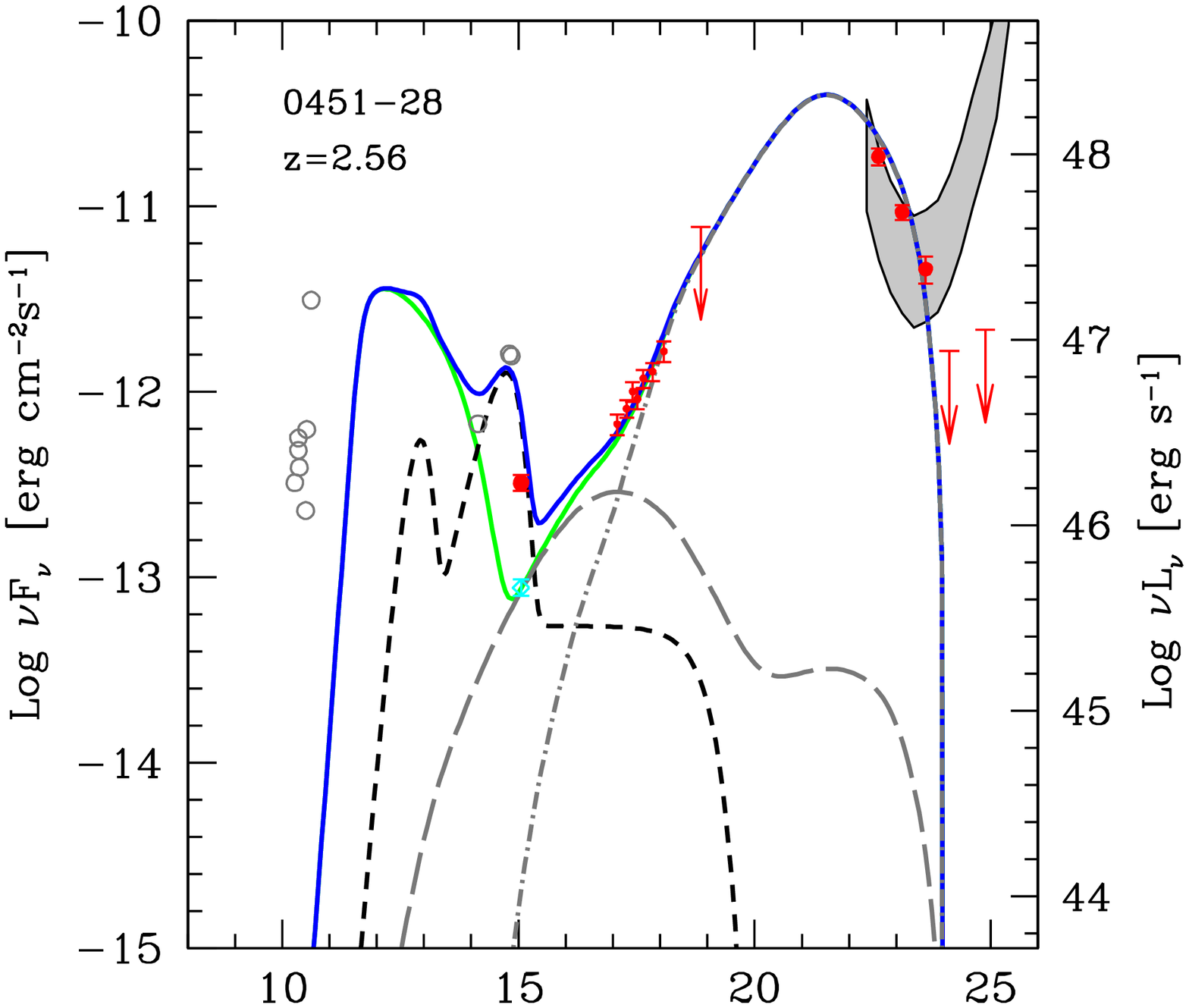,width=9cm,height=7cm}
\vskip -1.3 cm \hskip -0.4 cm
\psfig{figure=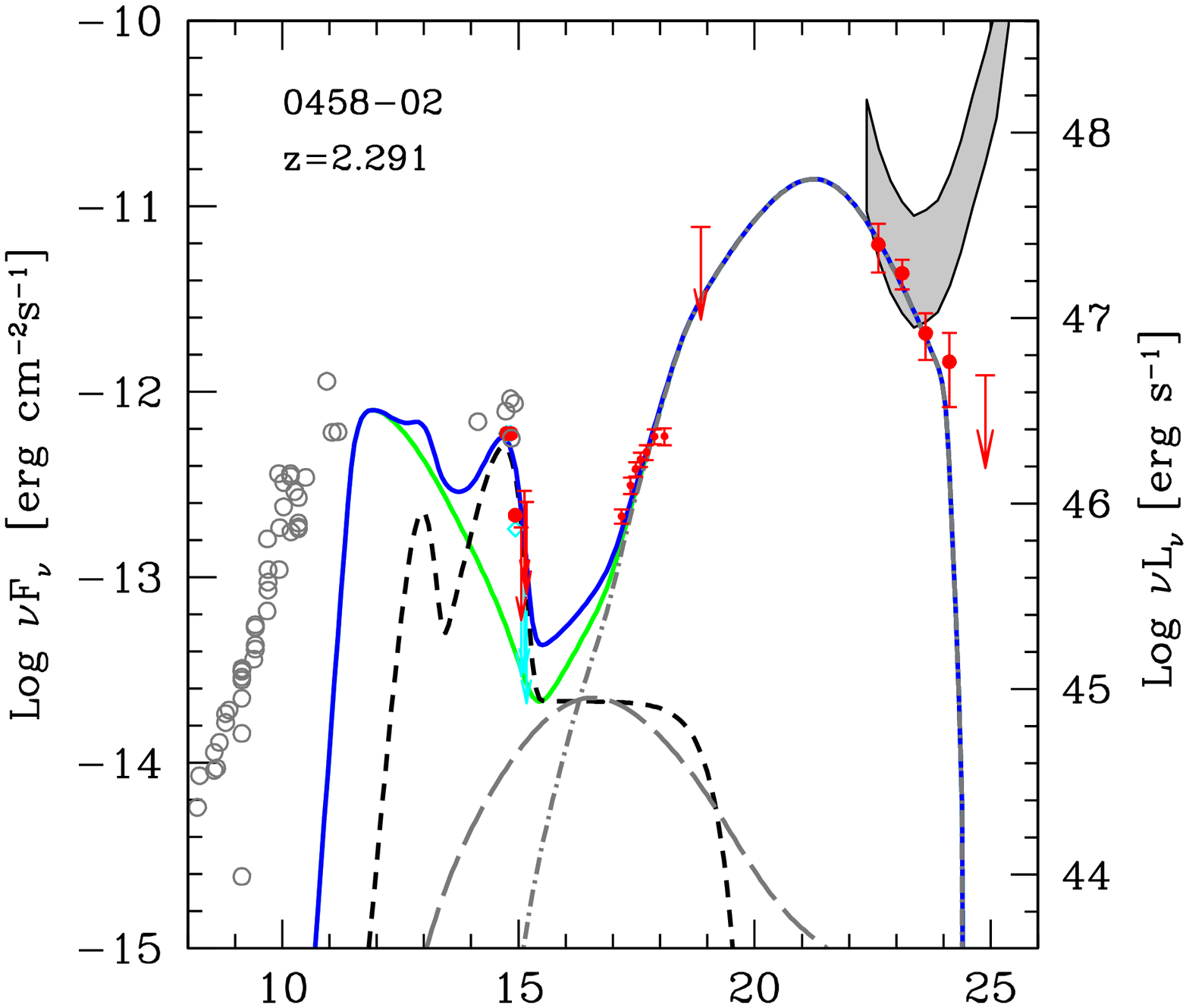,width=9cm,height=7cm}
\vskip -1.3 cm \hskip -0.4 cm
\psfig{figure=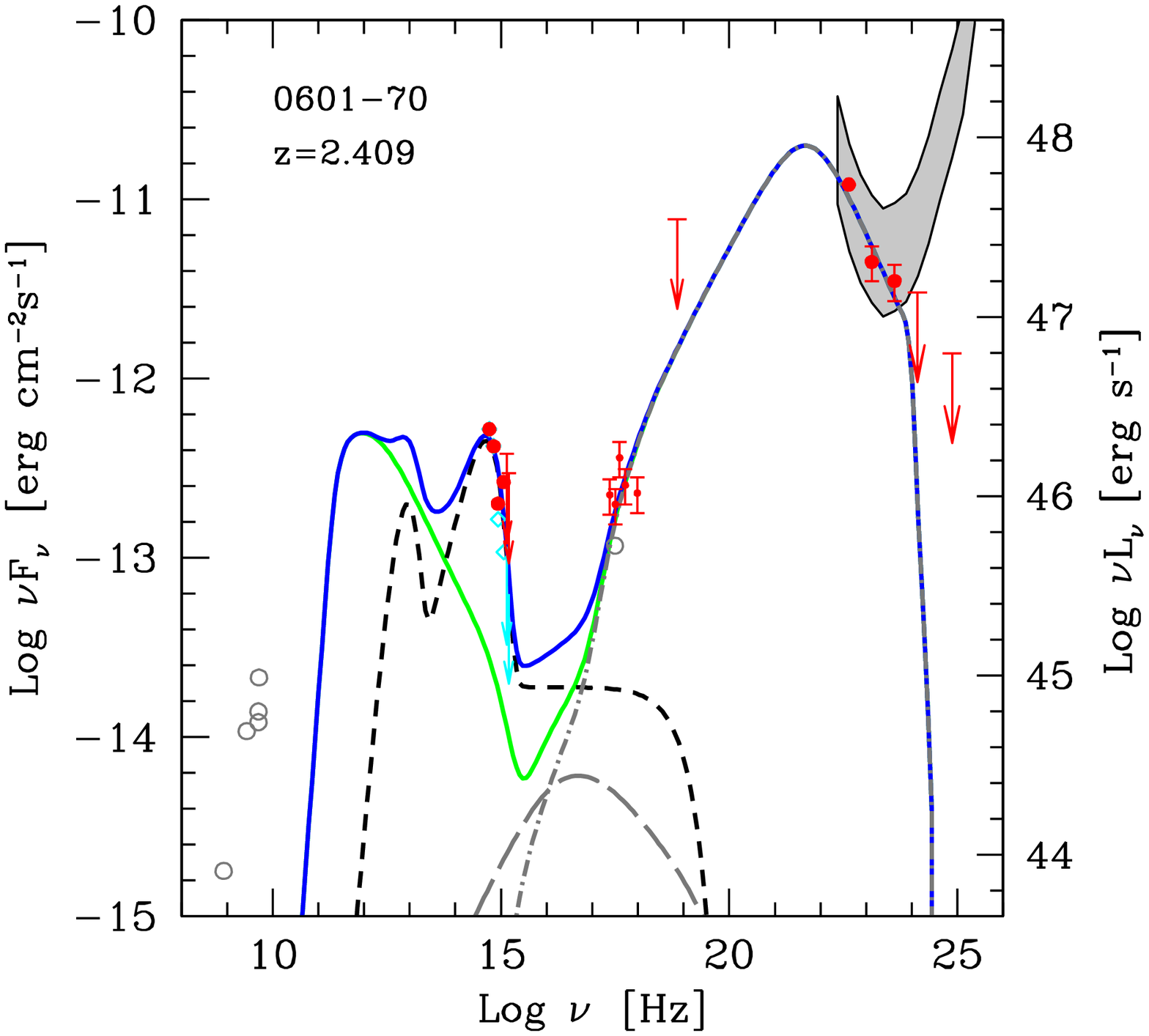,width=9cm,height=6.9cm}
\vskip -0.8 cm
\caption{
SED of PKS 0420+022, PKS 0451--28, PKS 0458--02 and PKS 0601--70.
Symbols and lines as in Fig. \ref{f1}.
}
\label{f2}
\end{figure}

\begin{figure}
\vskip -0.6 cm \hskip -0.4 cm
\psfig{figure=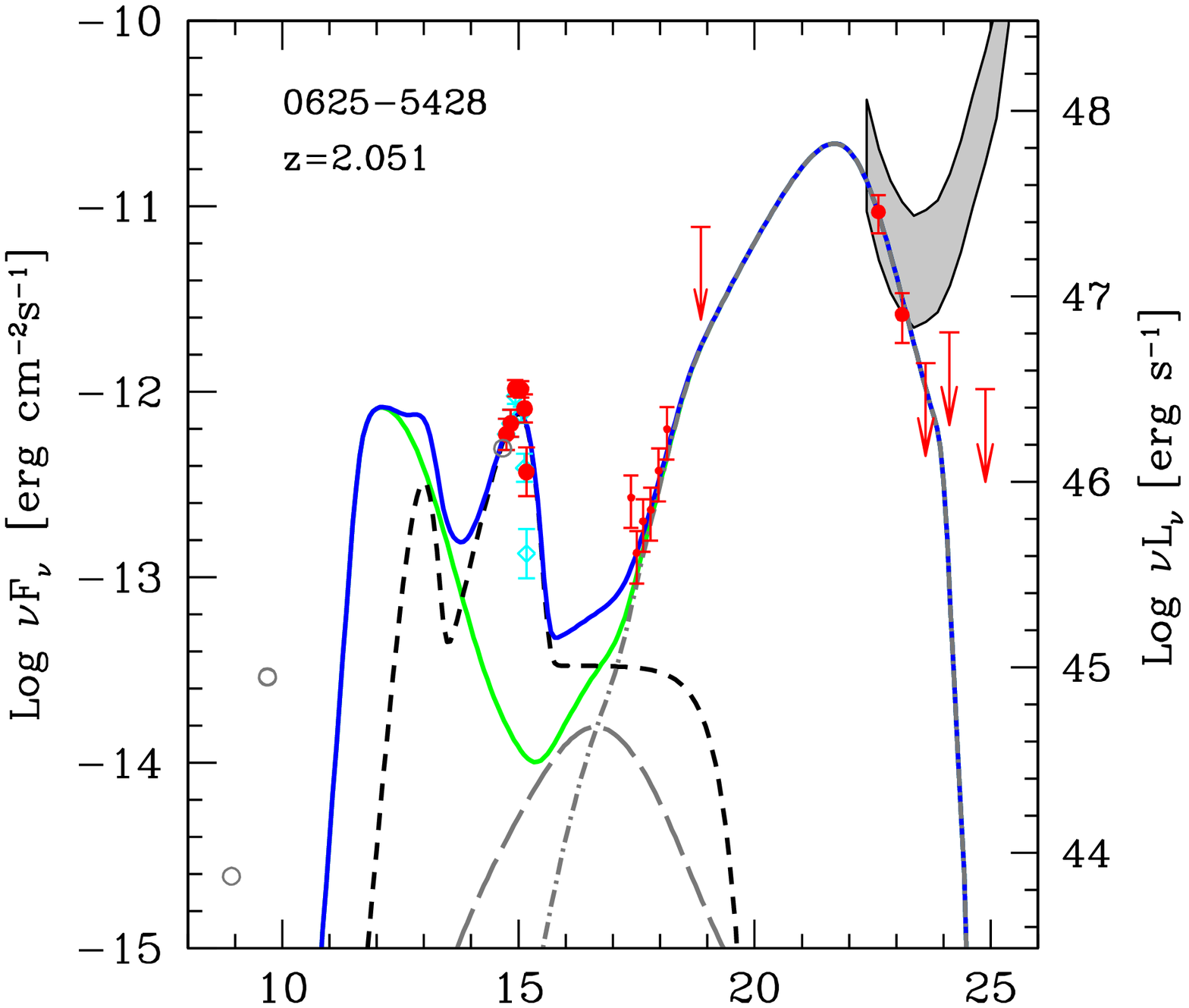,width=9cm,height=7cm}
\vskip -1.3 cm \hskip -0.4 cm
\psfig{figure=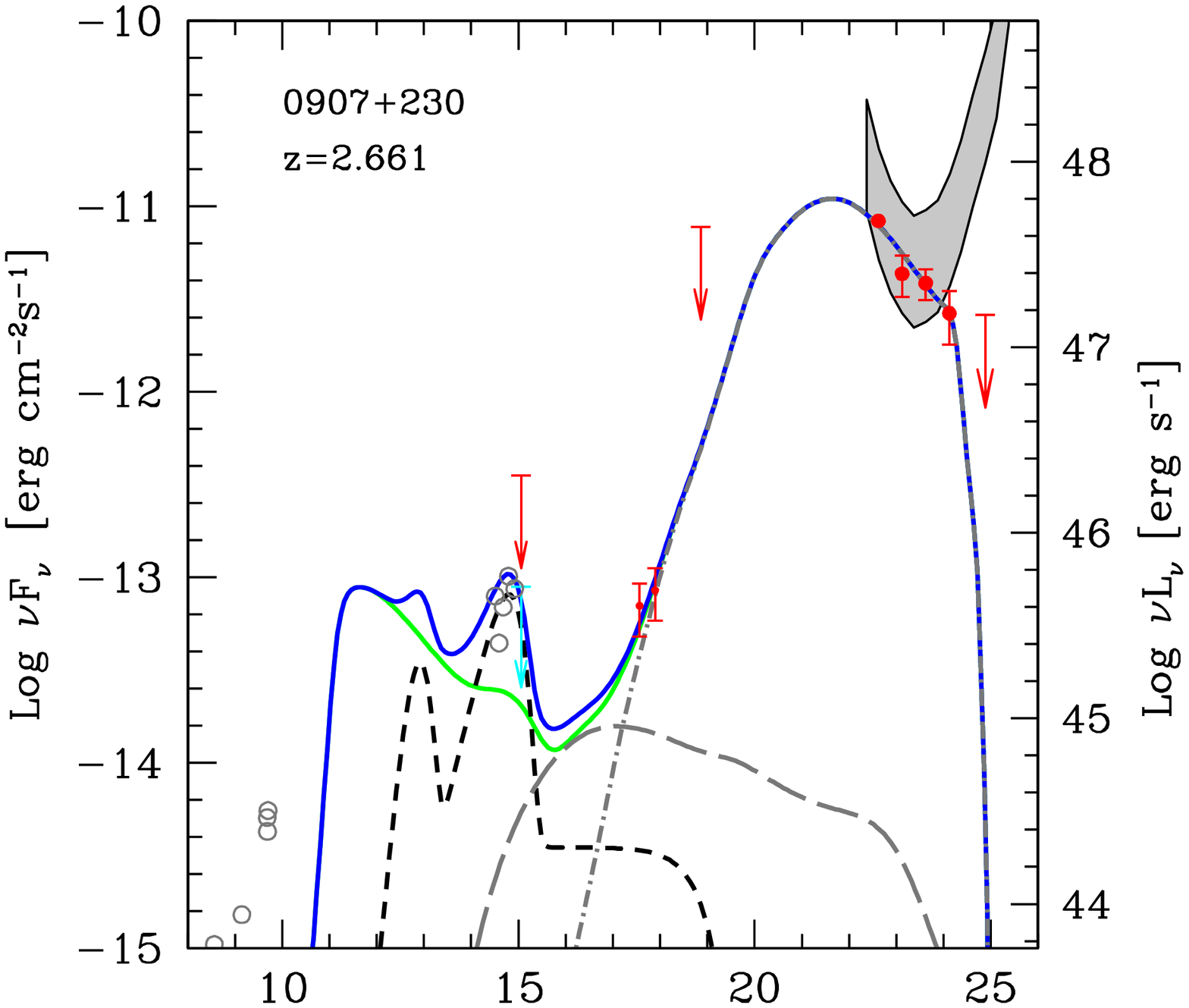,width=9cm,height=7cm}
\vskip -1.3 cm \hskip -0.4 cm
\psfig{figure=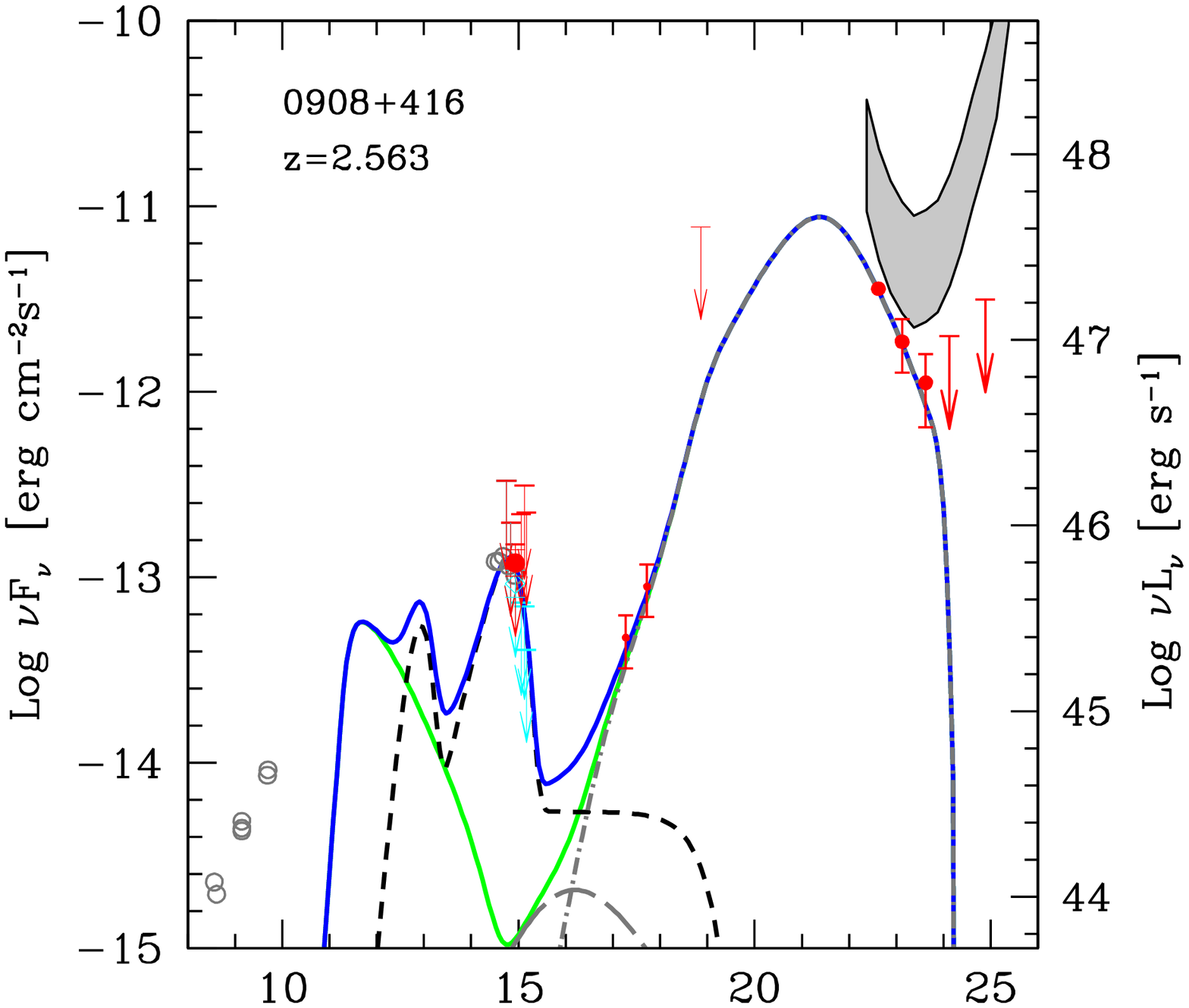,width=9cm,height=7cm}
\vskip -1.3 cm \hskip -0.4 cm
\psfig{figure=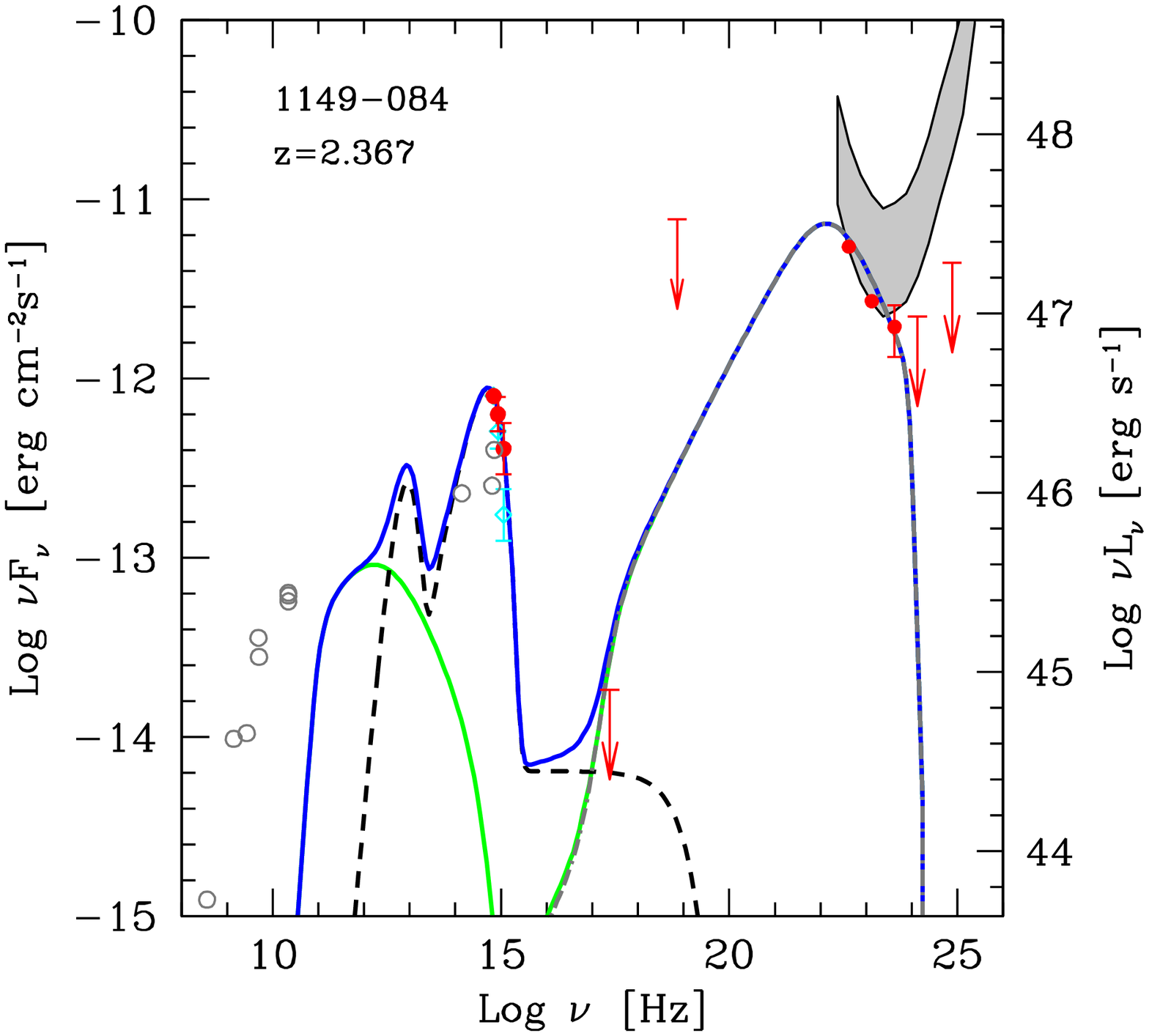,width=9cm,height=7cm}
\vskip -0.8 cm
\caption{
SED of PMN 0625--5438, TXS 0907+230, TXS 0908+416 and PKS 1149--084. 
Symbols and lines as in Fig. \ref{f1}.
}
\label{f3}
\end{figure}

\begin{figure}
\vskip -0.6 cm \hskip -0.4 cm
\psfig{figure=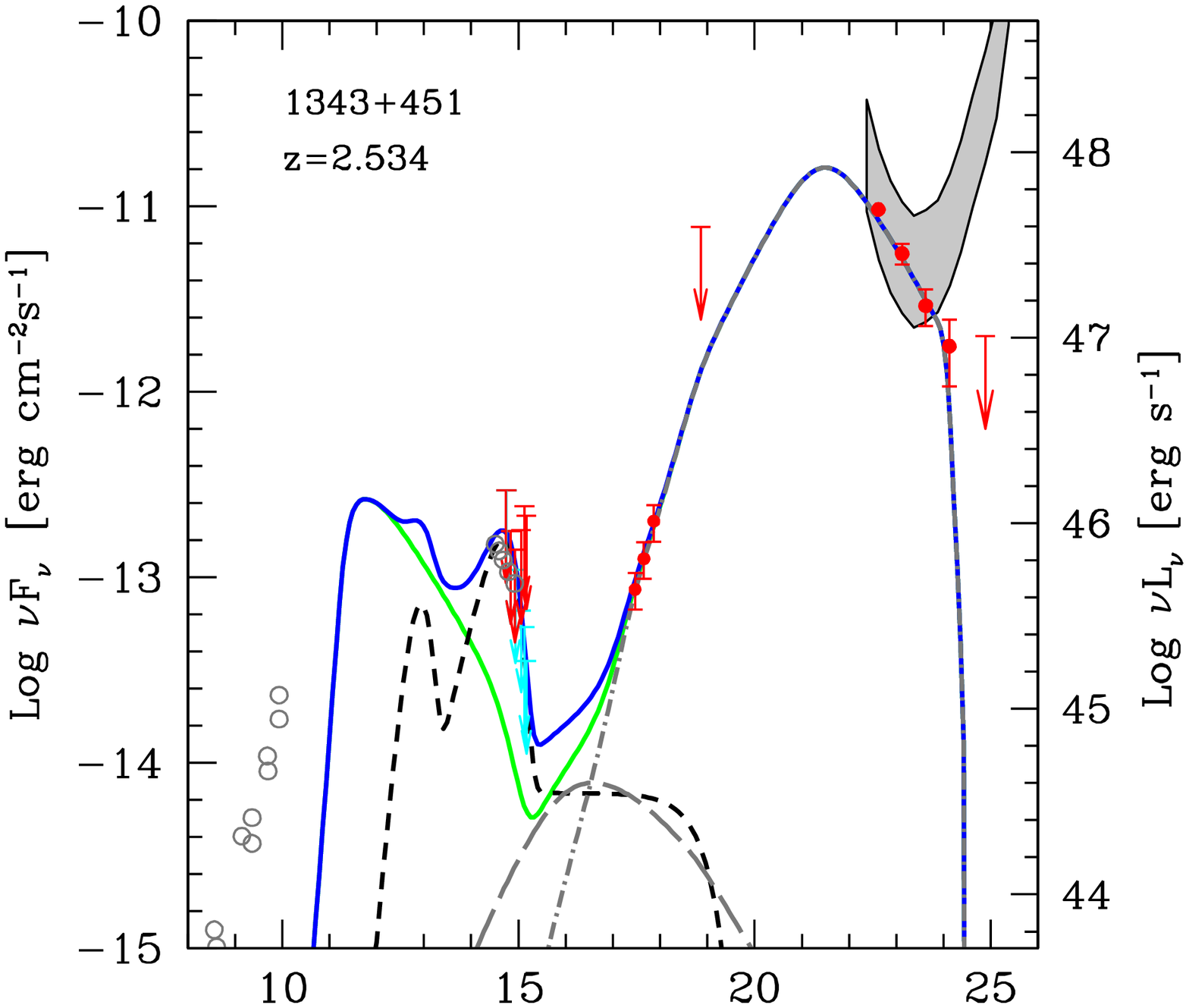,width=9cm,height=7cm}
\vskip -1.3 cm \hskip -0.4 cm
\psfig{figure=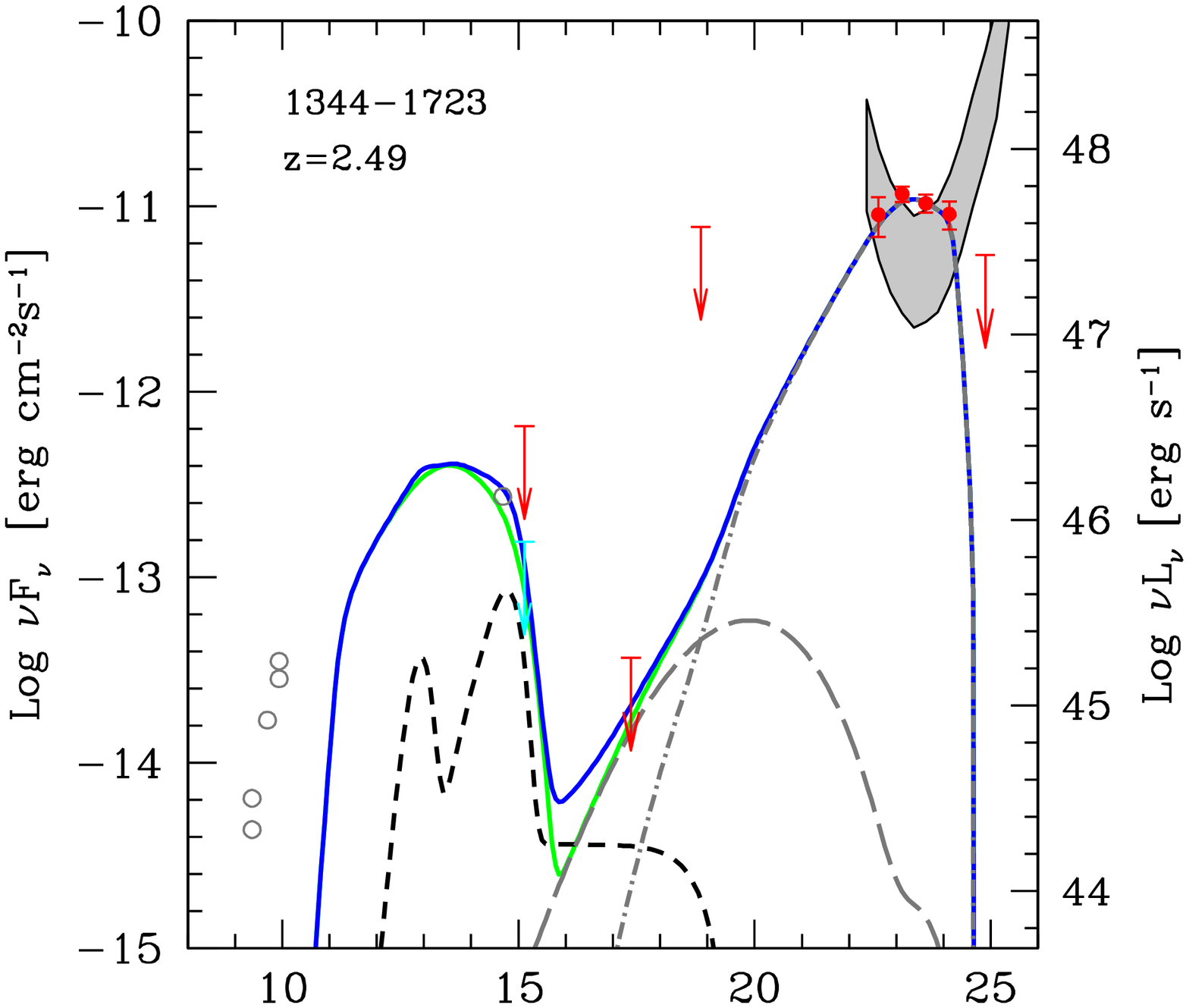,width=9cm,height=7cm}
\vskip -1.3 cm \hskip -0.4 cm
\psfig{figure=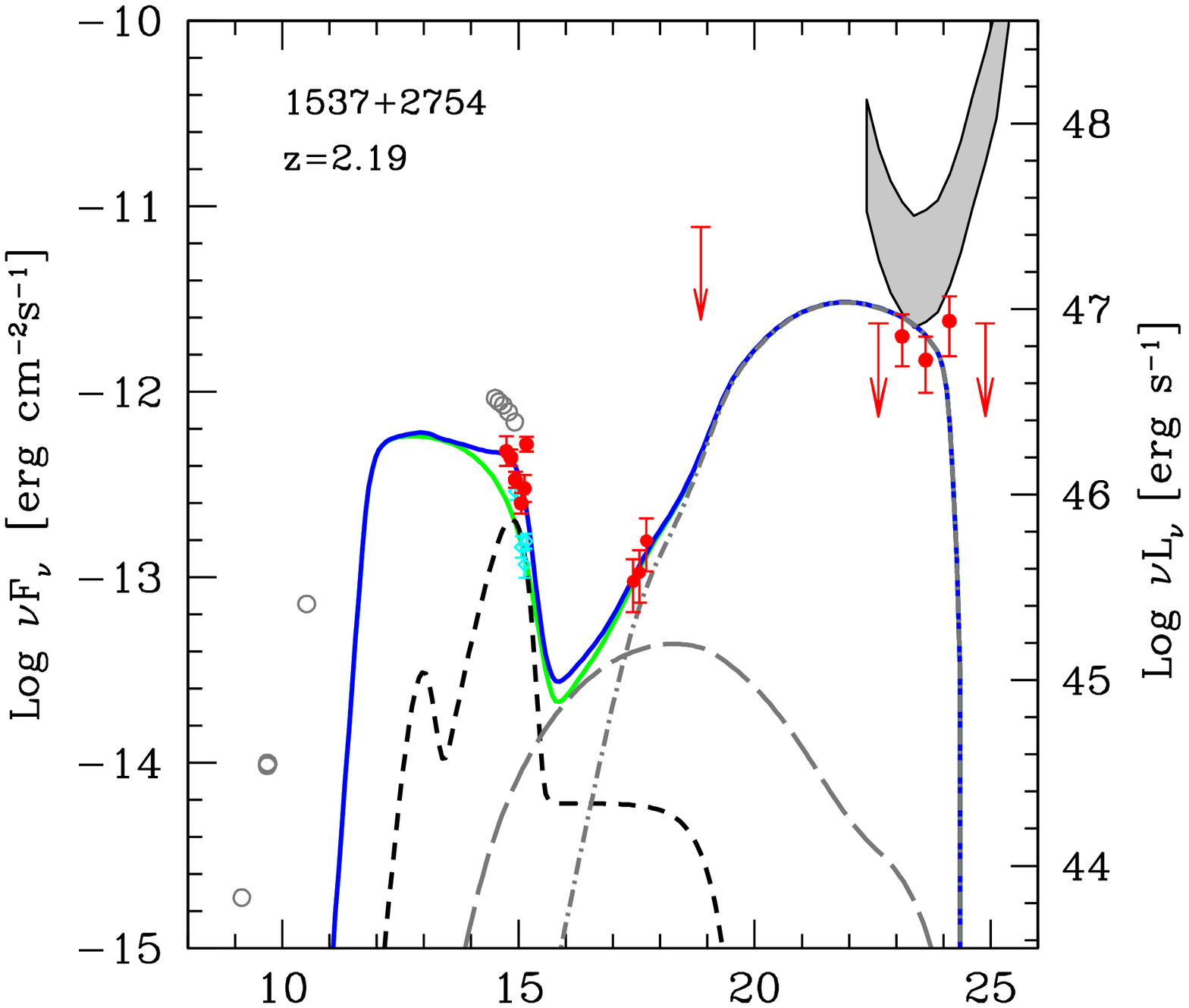,width=9cm,height=7cm}
\vskip -1.3 cm \hskip -0.4 cm
\psfig{figure=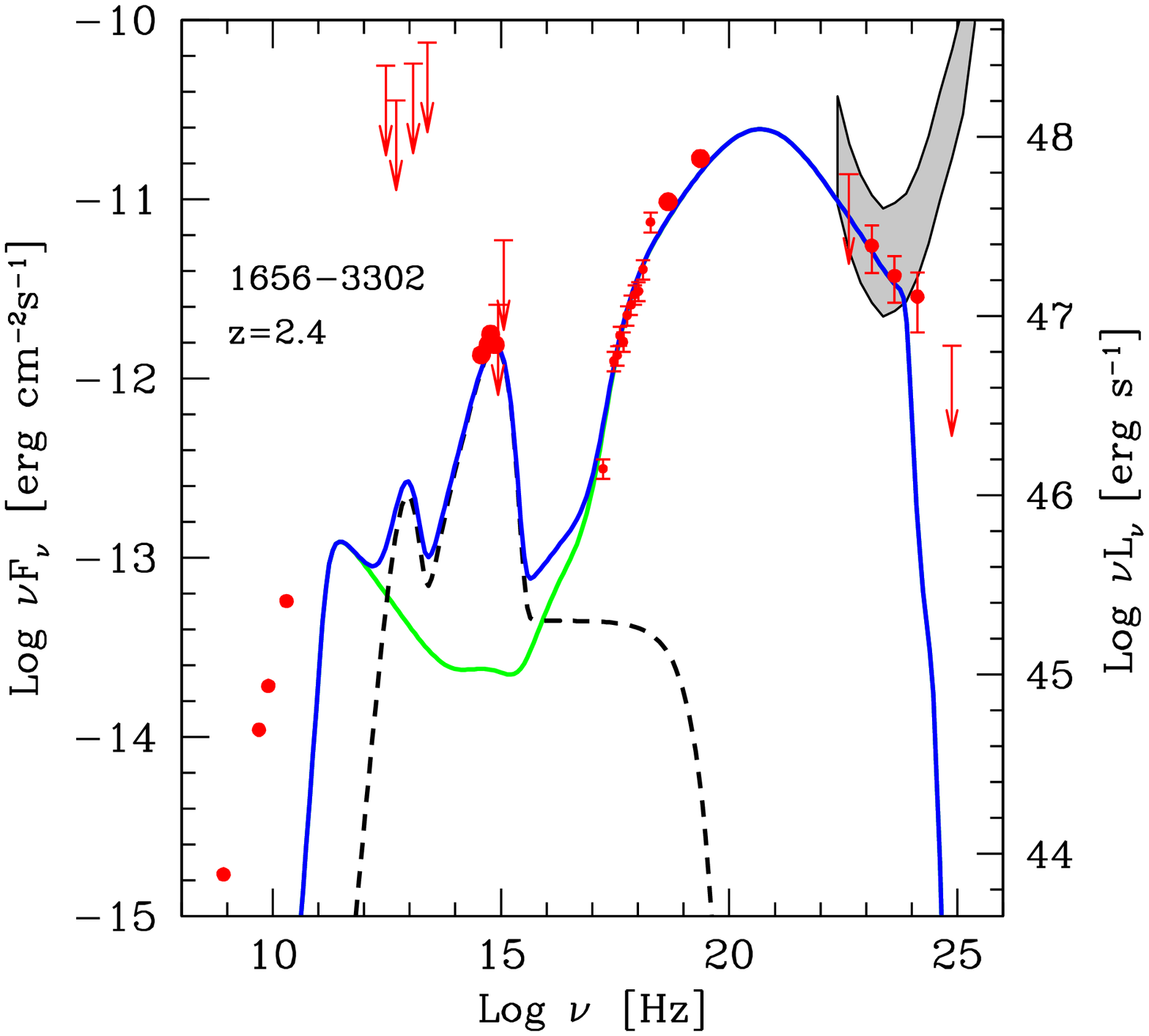,width=9cm,height=7cm}
\vskip -0.8 cm
\caption{
SED of TXS 1343+451, PMN 1344--1723, [WB92] 1537+2754 and 
SWIFT J1656--3302.
Symbols and lines as in Fig. \ref{f1}.
}
\label{f4}
\end{figure}

\begin{figure}
\vskip -0.6 cm \hskip -0.4 cm
\psfig{figure=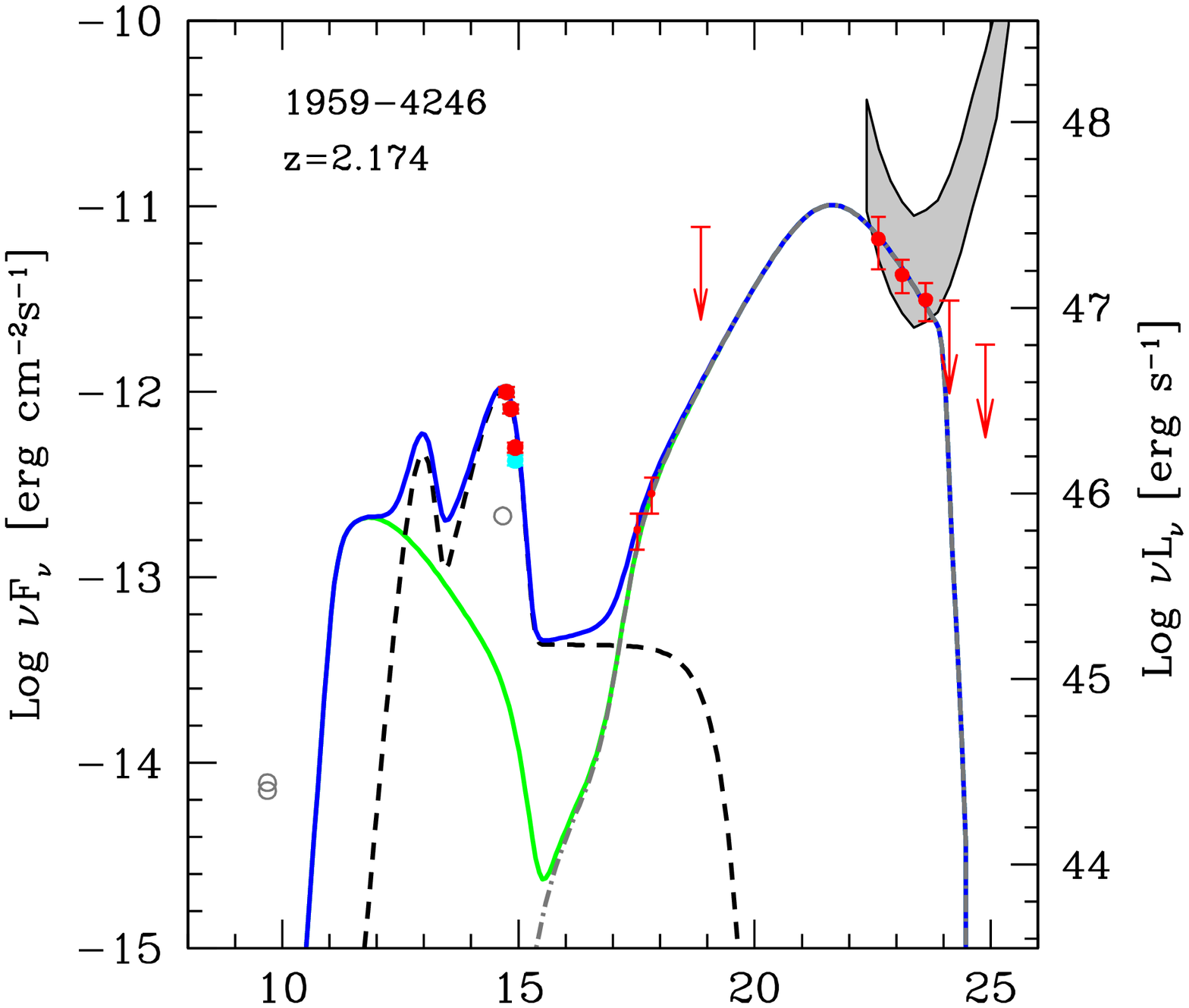,width=9cm,height=7cm}
\vskip -1.3 cm \hskip -0.4 cm
\psfig{figure=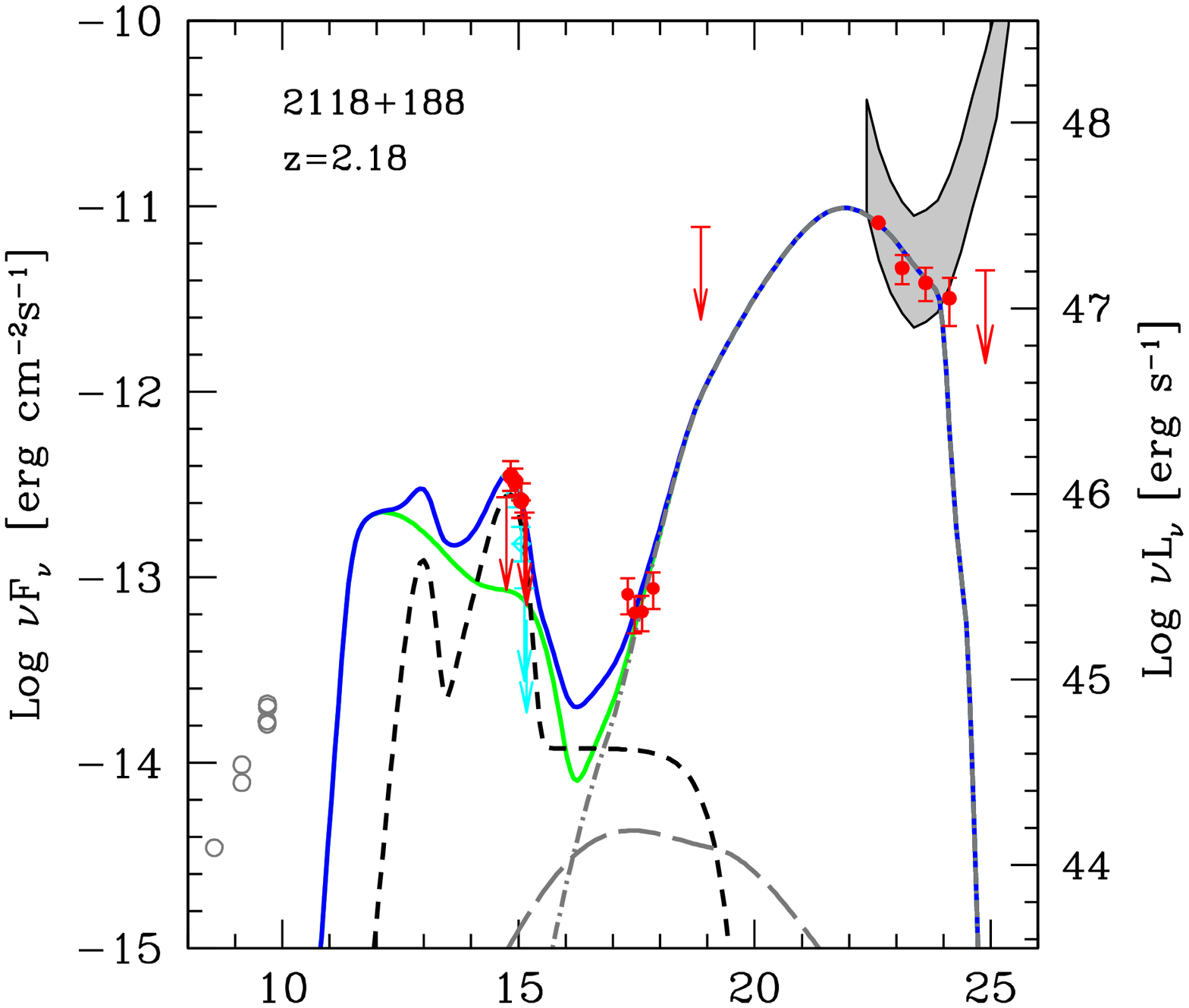,width=9cm,height=7cm}
\vskip -1.3 cm \hskip -0.4 cm
\psfig{figure=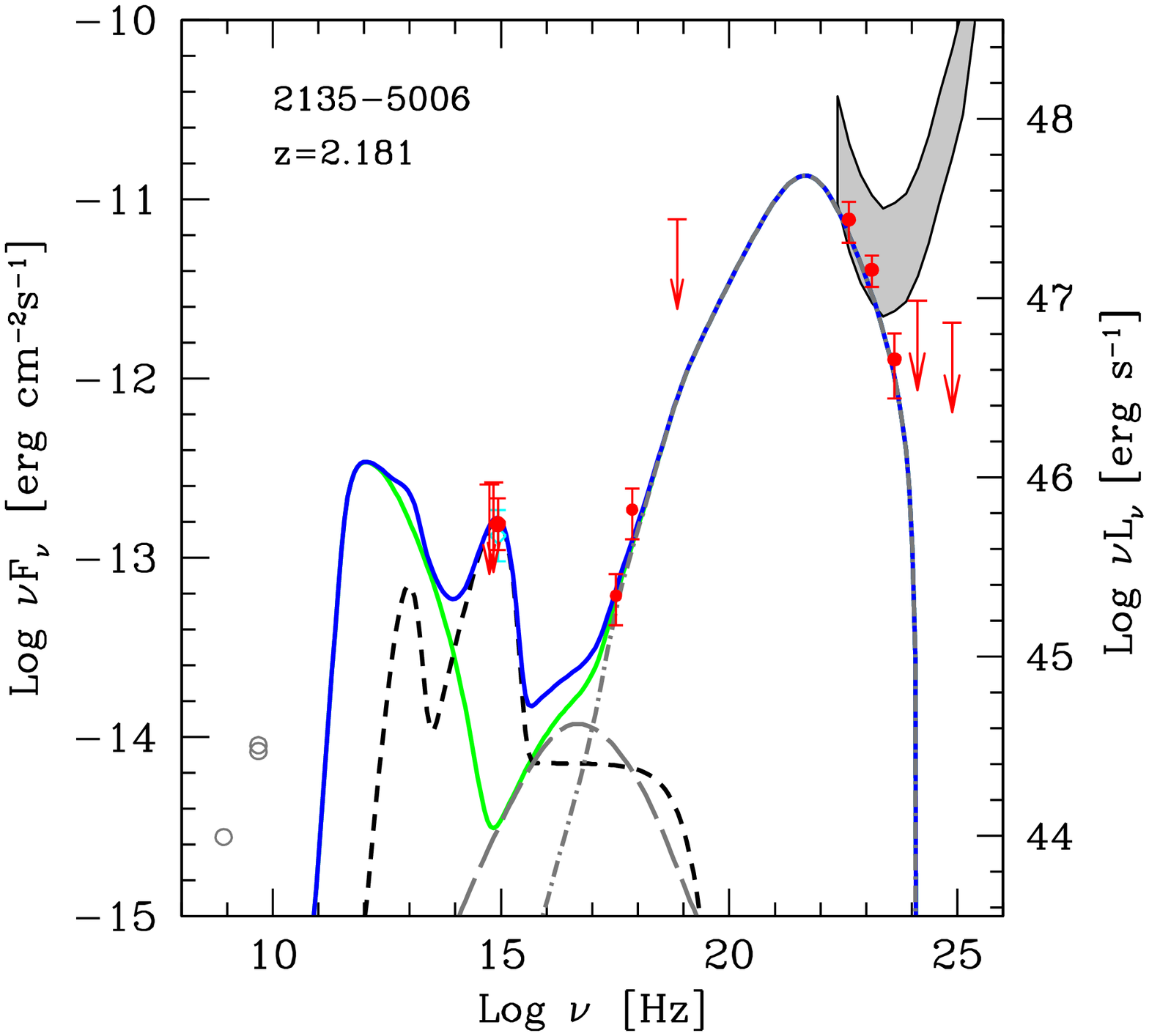,width=9cm,height=7cm}
\vskip -0.8 cm
\caption{
SED of PMN 1959--4246, TXS 2118+188 and PMN 2135--5006
Symbols and lines as in Fig. \ref{f1}.
}
\label{f5}
\end{figure}

\section{Modelling the SED}
\label{modelling}

To model the SEDs of the blazars in this sample we used the same model
used in G10. It is a one--zone, leptonic
model, fully discussed in Ghisellini \& Tavecchio (2009).
In that paper we emphasize the relative importance of
the different sources of the seed photons for the inverse Compton
scattering process, and how they change as a function of the 
distance of the emitting region from the black hole.
Here we briefly summarize the main characteristics of the model.

The source is assumed spherical (radius $R$) and located at a distance
$R_{\rm diss}$ from the central black hole. 
The emitting electrons are injected at
a rate $Q(\gamma)$ [cm$^{-3}$ s$^{-1}$] for a finite time equal to the 
light crossing time $R/c$. 
The shape of $Q(\gamma)$ we adopt is assumed to be a smoothly broken power law
with a break at $\gamma_{\rm b}$:
\begin{equation}
Q(\gamma)  \, = \, Q_0\, { (\gamma/\gamma_{\rm b})^{-s_1} \over 1+
(\gamma/\gamma_{\rm b})^{-s_1+s_2} }
\label{qgamma}
\end{equation}

The emitting region is moving with a  velocity $\beta c$
corresponding to a bulk Lorentz factor $\Gamma$.
We observe the source at the viewing angle $\theta_{\rm v}$ and the Doppler
factor is $\delta = 1/[\Gamma(1-\beta\cos\theta_{\rm v})]$.
The magnetic field $B$ is tangled and uniform throughout the emitting region.
We take into account several sources of radiation externally to the jet:
i) the broad line photons, assumed to re--emit 10\% of the accretion luminosity
from a shell--like distribution of clouds located at a distance 
$R_{\rm BLR}=10^{17}L_{\rm d, 45}^{1/2}$ cm;
ii) the IR emission from a dusty torus, located at a distance
$R_{\rm IR}=2.5\times 10^{18}L_{\rm d, 45}^{1/2}$ cm;
iii) the direct emission from the accretion disk, including its X--ray corona.
Furthermore we take into account 
the starlight contribution from the inner region of the host galaxy
and the cosmic background radiation, but these photon sources
are unimportant in our case.
All these contributions are evaluated in the blob comoving frame, where we 
calculate the corresponding inverse Compton radiation from all these components, 
and then transform into the observer frame.

We calculate the energy distribution $N(\gamma)$ [cm$^{-3}$]
of the emitting particles at the particular time $R/c$, 
when the injection process ends. 
Our numerical code solves the continuity equation which includes injection, 
radiative cooling and $e^\pm$ pair production and reprocessing. 
Ours is not a time dependent code: we give a ``snapshot" of the 
predicted SED at the time $R/c$, when the particle distribution $N(\gamma)$ 
and consequently the produced flux are at their maximum.

For all sources in our sample, the radiative cooling time of the 
particles is short, shorter than $R/c$ even for low energetic particles.
In Tab. \ref{para} (last column) we have listed the values of $\gamma_{\rm c}$,
that is the minimum value of the random Lorentz factor 
of electrons cooling in one light crossing time.
Since it is always smaller than $\gamma_{\rm b}$, almost all
the energy injected in the form of relativistic electrons is radiated away.
Most of the cooling is due to the inverse Compton scattering with broad line photons,
with a minor contribution from the synchrotron and the self--Compton process. 
Therefore we always are in the {\it fast cooling} regime 
(i.e. $\gamma_{\rm c}<\gamma_{\rm b}$).
In this regime the produced luminosity {\it does not depend} on the amount of 
the radiation energy density, but only on the energy content of the 
injected relativistic electrons.

Another implication is that, 
at lower energies, the $N(\gamma)$ distribution is proportional
to $\gamma^{-2}$, while, above $\gamma_{\rm b}$, $N(\gamma)\propto \gamma^{-(s_2+1)}$.
The electrons emitting most of the observed radiation have energies 
$\gamma_{\rm peak}$ which is close to $\gamma_{\rm b}$ 
(but these two energies are not exactly equal, due to the curved 
injected spectrum).

The accretion disk component is calculated assuming a standard
optically thick geometrically thin Shakura \& Sunjaev (1973) disk.
The emission is locally a black body.
The temperature profile of the disk is given e.g. in Frank, King \& Raine (2002).
Since the optical--UV emission
is the sum of the accretion disk and the jet non--thermal
components, for a few sources there is some degeneracy when deriving the 
black hole mass and the accretion rate.

We model at the same time the thermal disk (and IR torus) radiation 
and the non--thermal jet--emission.
The link between these two components is given by the amount of 
radiation energy density (as seen in the comoving frame of the emitting
blob) coming directly from the accretion disk or reprocessed by the BLR and
the IR torus.
This radiation energy density depends mainly on $R_{\rm diss}$, but
not on the adopted accretion rate or black hole mass
(they are in any case chosen to reproduce the observed thermal disk 
luminosity).

By estimating the physical parameters of the source we can calculate
the power that the jet carries  in the form of radiation
($P_{\rm r}$), magnetic field ($P_{\rm B}$), relativistic electrons
($P_{\rm e}$) and cold protons ($P_{\rm p}$) assuming one proton per
electron.
These powers are calculated according to:
\begin{equation}
P_{\rm i} \, =\, \pi R^2 \Gamma^2 c U^\prime_{\rm i}
\label{power}
\end{equation}
where $U^\prime$ is the energy density of the $i_{\rm th}$ component
in the comoving frame.

\subsection{Intervening Lyman--$\alpha$ absorption}

Being at $z>2$ the optical--UV flux of the blazars in our sample 
could be affected by absorption of neutral hydrogen in intervening 
Lyman--$\alpha$ absorption systems. 
To correct for this, we use the attenuation calculated in G10
specifically for the UVOT filters, illustrated in Fig. 3 of that paper.

Full details of our calculation will be described in Haardt et al. 
(in preparation), together with a more refined treatment of the mean 
attenuation and its variance around the mean.
The current procedure is very crude, especially when the attenuation is large
(i.e. optical depths larger than unity) because in such cases most of the 
attenuation is due to very few clouds, implying a large variance.
However, we note that the variance of the attenuation is largely 
reduced when the actual filter width is taken into account (Madau 1995).
Our absorption model results in a mean number of thick systems which is
$<1$ for $z\lsim 4$, so we do not expect excessive off--set of the 
attenuation along individual line of sight with respect to the mean value.

When presenting the SED of our sources, we will show 
both the fluxes and upper limits de--reddened for the extinction 
due to our Galaxy and the fluxes (and upper limits) obtained
by de--absorbing them with the $\tau_{\rm eff}$ shown in Fig. 3 in G10.

\begin{figure}
\vskip -0.8 cm 
\psfig{figure=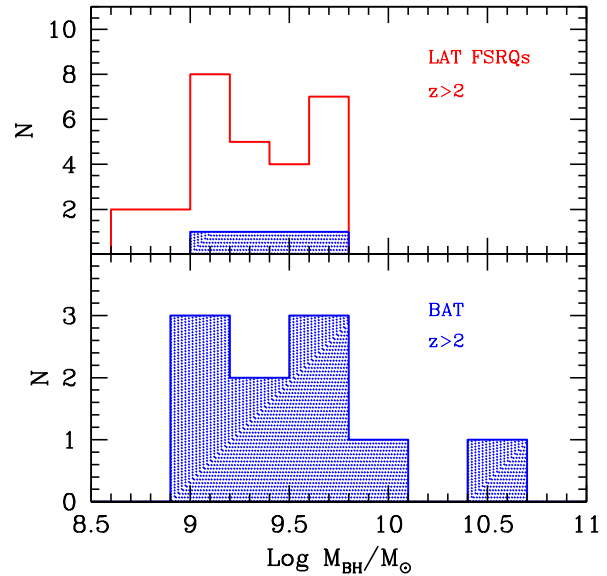,width=9cm,height=9cm}
\vskip -0.4 cm
\caption{Distribution of the black hole masses 
derived for the $z>2$ {\it Fermi}/LAT (top) 
and {\it Swift}/BAT samples (bottom).
The hatched area in the top panel corresponds to the 4 blazars
in common.
}
\label{masses}
\end{figure}

\begin{table*} 
\centering
\begin{tabular}{llllllllllllll}
\hline
\hline
Name   &$z$ &$R_{\rm diss}$ &$M$ &$R_{\rm BLR}$ &$P^\prime_{\rm i}$ &$L_{\rm d}$ &$B$ &$\Gamma$ 
    &$\gamma_{\rm b}$ &$\gamma_{\rm max}$ &$s_1$  &$s_2$  &$\gamma_{\rm c}$  \\
~[1]      &[2] &[3] &[4] &[5] &[6] &[7] &[8] &[9] &[10] &[11]  &[12] &[13]  &[14]\\
\hline   
0106+01    &2.107 &900 (600)   &5e9   &866   &0.08  &75 (0.1)   &1.13 &14   &300 &5e3   &0    &3.1  &2.0 \\ 
0157--4614 &2.287 &195 (1.3e3) &5e8   &274   &0.015 &7.5 (0.1)  &1.54 &15   &200 &2e3   &--1  &3.0  &5.7   \\
0242+23    &2.243 &420 (700)   &2e9   &812   &0.022 &66 (0.22)  &2.13 &15   &220 &2e3   &0.5  &3.1  &2.6  \\    
0322+222   &2.066 &450 (500)   &3e9   &671   &0.06  &45 (0.1)   &2.06 &12   &150 &3e3   &0.5  &3.1  &3.7   \\
0420+022   &2.277 &210 (1.4e3) &5e8   &725   &0.02  &52.5 (0.7) &3.79 &15   &300 &2e3   &--1  &3.2  &4.9   \\    
0451--28   &2.56  &540 (450)   &4e9   &1.1e3 &0.24  &120 (0.2)  &2.66 &10   &180 &2e3   &0.   &2.6  &4.1  \\ 
0458--02   &2.291 &472 (450)   &3.5e9 &606   &0.07  &37 (0.07)  &2.14 &10   &200 &5e3   &0.8  &3.0  &4.9  \\ 
0601--70   &2.409 &525 (500)   &3.5e9 &606   &0.05  &37 (0.07)  &1.83 &12.9 &190 &5e3   &--1  &3.1  &2.8  \\ 
0625--5438 &2.051 &270 (900)   &1e9   &648   &0.03  &42 (0.28)  &2.64 &15   &240 &5e3   &0    &4.0  &3.9   \\    
0907+230   &2.661 &360 (1.5e3) &8e8   &290   &0.05  &8.4 (0.07) &0.36 &13   &300 &1.7e4 &0.75 &2.8  &32.7   \\ 
0908+416   &2.563 &180 (600)   &1e9   &346   &0.025 &12 (0.08)  &1.06 &14   &150 &3e3   &0    &3.1  &7.0   \\ 
1149--084  &2.367 &720 (600)   &4e9   &849   &0.015 &72 (0.12)  &1.39 &14   &300 &3e3   &--1  &3.0  &1.8   \\ 
1343+451   &2.534 &420 (700)   &2e9   &387   &0.045 &15 (0.05)  &1.09 &14   &150 &5e3   &--1  &2.8  &6.5  \\ 
1344--1723 &2.409 &330 (1.1e3) &1e9   &274   &0.027 &7.5 (0.05) &0.89 &13 &1.4e3 &8e3   &--1  &2.5  &26.2   \\    
1537+2754  &2.19  &120 (400)   &1e9   &367   &0.015 &13.5 (0.09)&4.42 &11.5 &60  &4e3   &0.5  &2.1  &12.2  \\ 
1656--3302 &2.4   &525 (700)   &2.5e9 &1.1e3 &0.07  &124 (0.33) &1.09 &15   &70  &1e4   &0.75 &2.85 &2.1 \\
1959--4246 &2.174 &825 (500)   &5.5e9 &812   &0.024 &66 (0.08)  &1.51 &12.9 &170 &5e3   &0    &2.7  &1.9  \\ 
2118+188   &2.18  &270 (600)   &1.5e9 &424   &0.022 &18 (0.08)  &1.85 &14   &250 &1e4   &0.5  &2.8  &4.6 \\    
2135--5006 &2.181 &189 (900)   &7e8   &324   &0.023 &10.5 (0.1) &2.02 &14   &180 &2e3   &--1  &3.2  &6.6  \\ 
\hline
0227--369  &2.115  &420 (700)  &2e9   &547   &0.08  &30 (0.1)   &1.5  &14   &200  &5e3  &0     &3.1  &3.0   \\
0347--211  &2.944  &750 (500)  &5e9   &866   &0.12  &75 (0.1)   &1.5  &12.9 &500  &3e3  &--1   &3.0  &1.8    \\
0528+134   &2.04   &420 (1400) &1e9   &866   &0.13  &75 (0.5)   &2.6  &13   &150  &3e3  &--1   &2.8  &3.3    \\
0537--286  &3.104  &420 (700)  &2e9   &735   &0.13  &54 (0.18)  &1.92 &15   &50   &2e3  &--1   &3    &2.6  \\ 
0743+259   &2.979  &1.65e3 (1.1e3)&5e9 &866  &0.24  &75 (0.1)   &0.1  &15   &200  &5e3  &0.75  &2.6  &102   \\
0805+614   &3.033  &270 (600)  &1.5e9 &581   &0.15  &34 (0.15)  &2.54 &14   &60   &3e3  &--0.5 &3    &4.3 \\  
0836+710   &2.172  &540 (600)  &3e9   &1.5e3 &0.22  &225 (0.5)  &3.28 &14   &90   &2e3  &--1   &3.6  &2.1  \\
0917--449  &2.1899 &900 (500)  &6e9   &1341  &0.1   &180 (0.2)  &1.95 &12.9 &50   &4e3  &--1   &2.6  &1.6   \\
1329--049  &2.15   &450 (1e3)  &1.5e9 &822   &0.07  &67.5 (0.3) &1.4  &15   &300  &5e3  &1     &3.3  &2.5  \\
\hline
\hline 
\end{tabular}
\vskip 0.4 true cm
\caption{List of parameters used to construct the theoretical SED.
Not all of them are ``input parameters" for the model, because $R_{\rm BLR}$
is uniquely determined from $L_{\rm d}$, and the cooling energy $\gamma_{\rm c}$ 
is a derived parameter.
Col. [1]: name;
Col. [2]: redshift;
Col. [3]: dissipation radius in units of $10^{15}$ cm and (in parenthesis) in units of \sc\ radii;
Col. [4]: black hole mass in solar masses;
Col. [5]: size of the BLR in units of $10^{15}$ cm;
Col. [6]: power injected in the blob calculated in the comoving frame, in units of $10^{45}$ erg s$^{-1}$; 
Col. [7]: accretion disk luminosity in units of $10^{45}$ erg s$^{-1}$ and
        (in parenthesis) in units of $L_{\rm Edd}$;
Col. [8]: magnetic field in Gauss;
Col. [9]: bulk Lorentz factor at $R_{\rm diss}$;
Col. [10] and [11]: break and maximum random Lorentz factors of the injected electrons;
Col. [12] and [13]: slopes of the injected electron distribution [$Q(\gamma)$] below and above $\gamma_{\rm b}$;
Col. [14] values of the minimum random Lorentz factor of those electrons cooling in one light crossing time.
The total X--ray corona luminosity is assumed to be in the range 10--30 per cent of $L_{\rm d}$.
Its spectral shape is assumed to be always $\propto \nu^{-1} \exp(-h\nu/150~{\rm keV})$.
The viewing angle $\theta_{\rm v}$ is $3^\circ$ for all sources.
}
\label{para}
\end{table*}

\section{Results}

Table \ref{para} lists all parameters used to model the SEDs of our blazars,
Table \ref{powers} lists the different forms of power carried by the jet
and Fig. \ref{f0}--\ref{f5} show the SEDs of the 19 blazars studied
in this paper and the corresponding fitting model.
In all figures we have marked with a grey shaded area the 
{\it Fermi}/LAT sensitivity, bounded on the bottom
by considering one year of operation and
a 5$\sigma$ detection level, and on the top by considering 3 months
and a 10$\sigma$ detection level (this assumes a common 
energy spectral index of $\alpha_\gamma\sim 1$, the sensitivity limit
for other spectral indices is slightly different, see Fig. 9 in A10).
All these sources were not detected in the first 3 months, in fact 
the (11 months) $\gamma$--ray data points are very close to the lower boundary
of the grey area.
There are exceptions: 
0106+01 (=4C+01.02) is brighter than the 3--months, 10$\sigma$
sensitivity limits even if it has not been included in LBAS.
This is due to a rather strong variability of the source,
fainter in the first 3 months and brighter soon after.
The same occurred for 1344--1723 and 0451--28.
The opposite happened for 0227--369, 0347--211 and 0528+134 
(i.e. they were brighter during the
first 3 months), but their flux, averaged over 11 months, was in any
case large enough to let their inclusion in the 1LAC sample.

Some of the sources have a sufficiently good IR--optical--UV
coverage to allow to see a peak of the SED in this band
(see for instance 
0420+022; 0451--28; 0458--02; 0625--5428; 0907+230;
0908+416; 1149--084; 1656--3302). 
The other sources have a SED consistent with a peak 
in this band, but the lack of data also allows for 
a peak at lower frequencies.
We interpret the peak in the optical band as due
to the accretion disk, and assume its presence
also in those blazars where it is allowed, but not
strictly required.
By assuming a standard Shakura--Sunyaev (1973) disk
we are then able to estimate both the black hole mass
and the accretion rate.
This important point has been discussed in G10,  in Ghisellini et al. (2009) 
(for S5 0014+813) and in Ghisellini \& Tavecchio (2009).

The radio data cannot be fitted by a simple one--zone model
specialized to fit the bulk of the emission, since the
latter must be emitted in a compact region, whose
radio flux is self--absorbed up to hundreds of GHz.
The radio emission should come from larger regions of the jet.
On the other hand, when possible, we try to have some
``continuity" between the non--thermal model continuum
and the radio fluxes (i.e. the model, in its low frequency
part, should not lie at too low or too high fluxes 
with respect to the radio data).
In the following we briefly comment on the
obtained parameters.

\vskip 0.3 cm                                                     
\noindent
{\it Dissipation region ---} The distance $R_{\rm diss}$ at which
most of the dissipation takes place is one of the key parameters
for the shape of the overall SED, since it controls the amount
of energy densities as seen in the comoving frame
(see Ghisellini \& Tavecchio 2009).
For almost all sources we have $R_{\rm diss} < R_{\rm BLR}$,
while for 0106+01, 0907+230, 1343+451, 1344--1723, 1959--4246,
$R_{\rm diss}$ is slightly larger than $R_{\rm BLR}$,
and for 0743+259 $R_{\rm diss}\sim 2 R_{\rm BLR}$.

In all sources the dominant cooling is trough inverse Compton off the 
seed photons of the BLR.
This is true also for the few blazars in which $R_{\rm diss}\ge R_{\rm BLR}$
since, even if $U^\prime_{\rm BLR}$ 
seen in the comoving frame is smaller, also $U_{\rm B}$ is smaller,
implying that the ratio $U^\prime_{\rm BLR}/U_B$ is similar to the values in other
sources [$U^\prime_{\rm BLR}/U_B$
ranges between $\sim$30 (1344--1723) and $\sim$100 (0106+01, 0907+230)].
However, the decreased cooling rate in 0743+259
makes $\gamma_{\rm c}$ to be larger ($\gamma_{\rm c} =102$, see Tab. \ref{para}).

With larger still $R_{\rm diss}\gg R_{\rm BLR}$, the main seed photons for the
Compton scattering process would became the photons produced by the the IR torus
(if it exists), but this case does not occur for our sources.

\vskip 0.3 cm
\noindent
{\it Compton dominance ---}
This is the ratio between the luminosity emitted at high frequencies
and the synchrotron luminosity.
The average magnetic field is found to be of the order of 1 Gauss,
with a corresponding magnetic energy density that is around two orders
of magnitude lower than the radiation energy density.
Correspondingly, all sources are Compton dominated.
 
\vskip 0.3 cm
\noindent
{\it Black hole masses ---} 
Fig. \ref{masses} shows the distribution of black hole masses for the
28 blazars at $z>2$ and compares them
with the distribution of masses for the high redshift BAT blazars.
Although the black hole masses of the BAT sample extend to larger values,
there are still too few sources to estimate if the two distributions
are different.
It is interesting to note that all but 3 sources
(0420+022, 0907+230 and 2135--5006) 
have black hole masses greater than $10^9M_\odot$. 
In Ghisellini, Tavecchio \& Ghirlanda  (2009) 
we considered the {\it Fermi} blazars of $\gamma$--ray
luminosity $L_\gamma>10^{48}$ erg s$^{-1}$, finding, with the same 
method and model applied here, that for all these sources the black
hole mass was greater than a billion solar masses.
Therefore {\it all} blazars with $L_\gamma>10^{48}$ erg s$^{-1}$
have black holes heavier than $10^9 M_\odot$, while the vast majority, but not all,
blazars at $z>2$ have such large black hole masses.
We searched in the literature other estimates of the black hole masses
for the objects in this sample, finding $M=2.3\times 10^9 M_\odot$
for 0836+710 (estimated by Liu, Jiang \& Gu 2006), and other few limits
for the black hole masses for 0836+710 and 0528+134, that were however
based assuming an isotropic $\gamma$--ray emission.

\begin{table} 
\centering
\begin{tabular}{lllll}
\hline
\hline
Name   &$\log P_{\rm r}$ &$\log P_{\rm B}$ &$\log P_{\rm e}$ &$\log P_{\rm p}$ \\
\hline   
0106+01      &46.19 &45.88 &44.71 &47.15   \\
0157--4614   &45.51 &44.88 &44.50 &46.71   \\
0242+23      &45.68 &45.83 &44.49 &46.96   \\ 
0322+222     &45.91 &45.67 &44.94 &47.36   \\ 
0420+022     &45.64 &45.73 &44.43 &46.68   \\    
0451--28     &46.35 &45.89 &45.22 &47.53   \\ 
0458--02     &45.81 &45.58 &44.92 &47.32   \\ 
0601--70     &45.91 &45.76 &44.60 &47.00   \\ 
0625--5438   &45.81 &45.63 &44.73 &47.03   \\    
0907+230     &45.86 &44.03 &45.30 &47.14   \\ 
0908+416     &45.66 &44.43 &44.80 &46.96   \\ 
1149--084    &45.47 &45.87 &43.83 &46.24   \\ 
1343+451     &45.93 &45.18 &44.94 &47.17   \\ 
1344--1723   &45.66 &44.74 &44.43 &46.03   \\    
1537+2754    &45.24 &45.14 &44.47 &46.58   \\ 
1656.3--3302 &46.17 &45.44 &45.23 &47.88   \\
1959--4246   &45.60 &46.06 &44.19 &46.67   \\ 
2118+188     &45.62 &45.26 &44.50 &46.81   \\    
2135--5006   &45.63 &45.03 &44.68 &46.93   \\ 
\hline
0227--369   &46.18 &45.49 &44.97 &47.34 \\ 
0347--211   &46.30 &45.91 &44.55 &46.92 \\ 
0528+134    &47.39 &45.86 &45.87 &48.31 \\    
0537-286    &46.43 &45.74 &45.52 &48.01 \\ 
0743+259    &46.53 &44.36 &46.28 &47.85 \\ 
0805+6144   &46.42 &45.34 &45.64 &47.99 \\
0836+710    &46.60 &46.36 &45.54 &48.00 \\
0917+449    &46.20 &46.29 &45.00 &47.57 \\
1329-049    &46.18 &45.53 &45.07 &47.65 \\
\hline
\hline 
\end{tabular}
\vskip 0.4 true cm
\caption{
Logarithm of the jet power in the form of radiation ($P_{\rm r}$), 
Poynting flux ($P_{\rm B}$),
bulk motion of electrons ($P_{\rm e}$) and protons ($P_{\rm p}$,
assuming one proton per emitting electron). 
Powers are in erg s$^{-1}$.
The bottom part of the table reports the data derived in
G10 and G09.
}
\label{powers}
\end{table}

\begin{figure}
\vskip -1.3 cm 
\hskip -2.8 cm
\psfig{figure=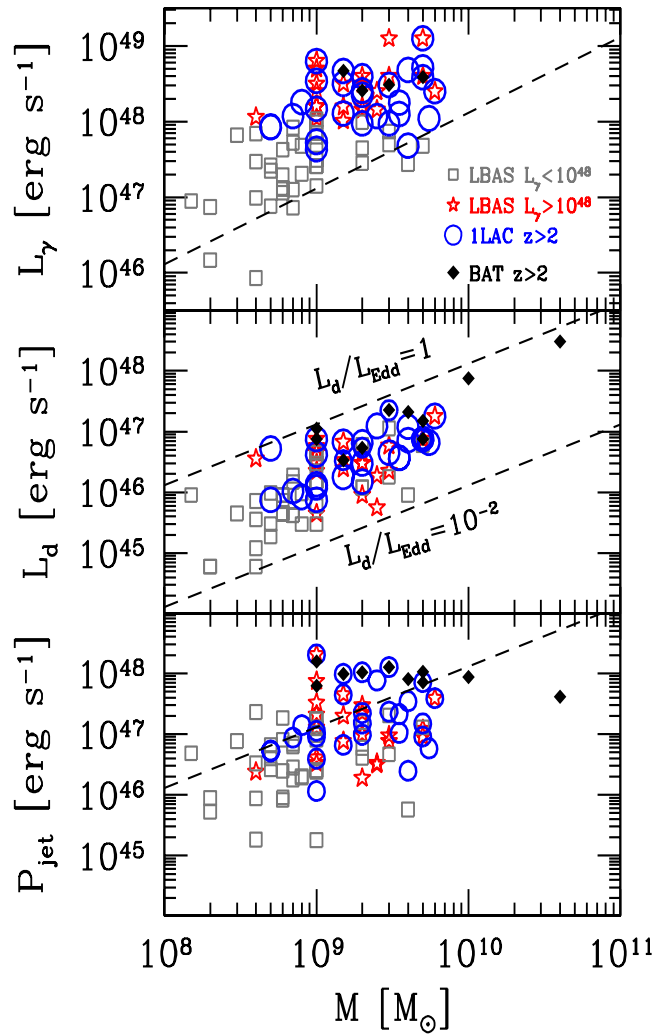,width=14.2cm,height=16.5cm}
\vskip -0.9 cm
\caption{
The observed $\gamma$--ray luminosity in the
0.1--10 GeV band, the accretion luminosity $L_{\rm d}$ 
and the total jet power $P_{\rm jet}$ as a function
of the derived black hole mass. 
All points correspond to FSRQs.
Different symbols correspond to LBAS FSRQs
with $L_\gamma>10^{48}$ erg s$^{-1}$ (red stars,
analysed in Ghisellini, Tavecchio \& Ghirlanda 2009);
1LAC FSRQs with $z>2$ (empty blue circles);
BAT FSRQs with $z>2$ (black diamonds, G10) and
the LBAS FSRQs with $L_\gamma<10^{48}$ erg s$^{-1}$ 
(grey squares, Ghisellini et al. 2010b).
The $\gamma$--ray luminosity (top panel) is the
observed beamed one, and it has not to be confused
with $P_{\rm r}$, the power spent by the jet to
produce the radiation we see.
The mid panel shows that all FSRQs have
disk luminosities between 0.01 and 1 Eddington luminosity.
The bottom panel shows that $P_{\rm jet}$ can be larger (but not by a
big factor) than the Eddington luminosity corresponding to the dashed line.
}
\label{powerm}
\end{figure}

\begin{figure}
\vskip -0.4 cm 
\psfig{figure=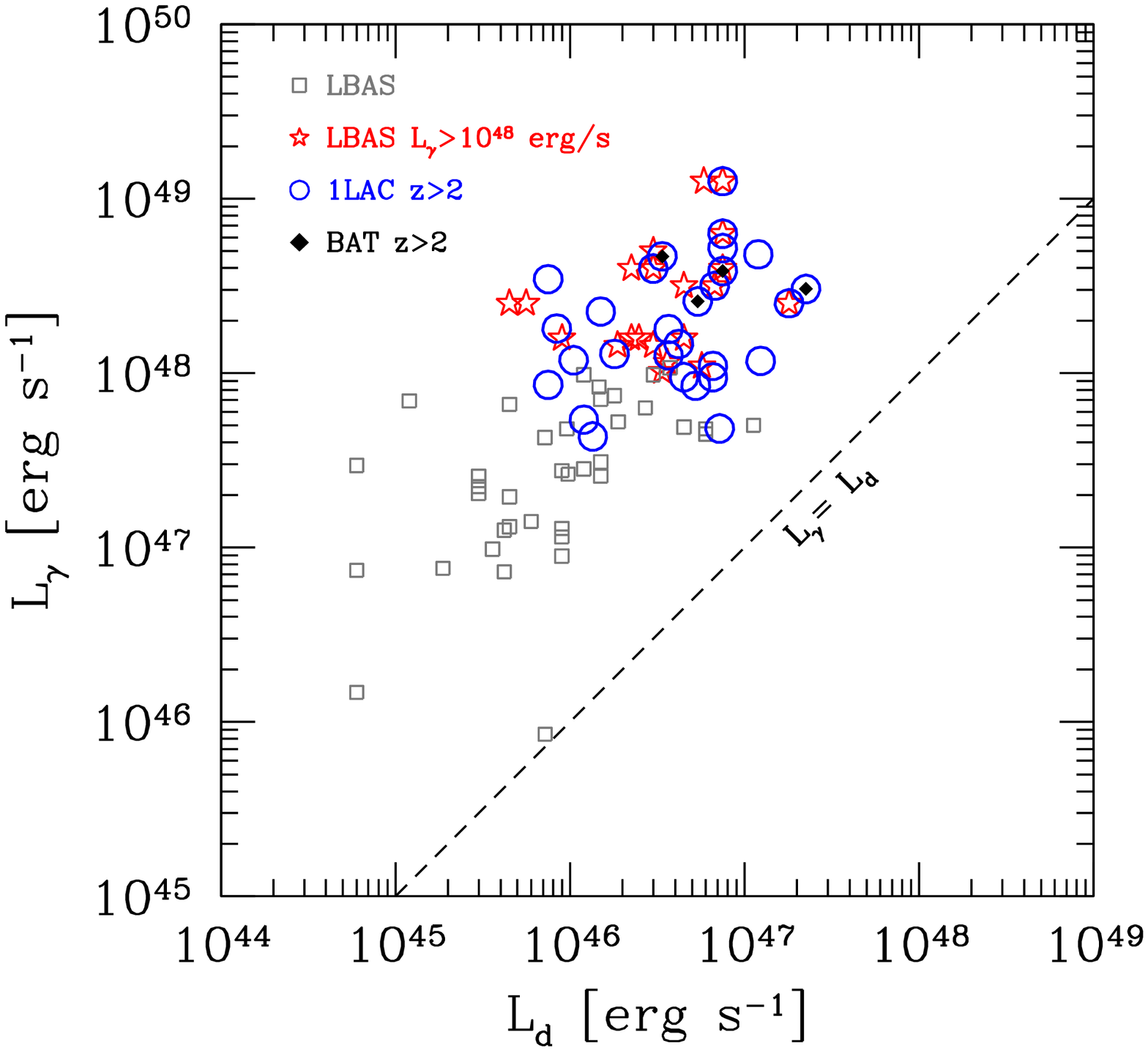,width=8cm,height=7.5cm}
\vskip -0.8 cm
\psfig{figure=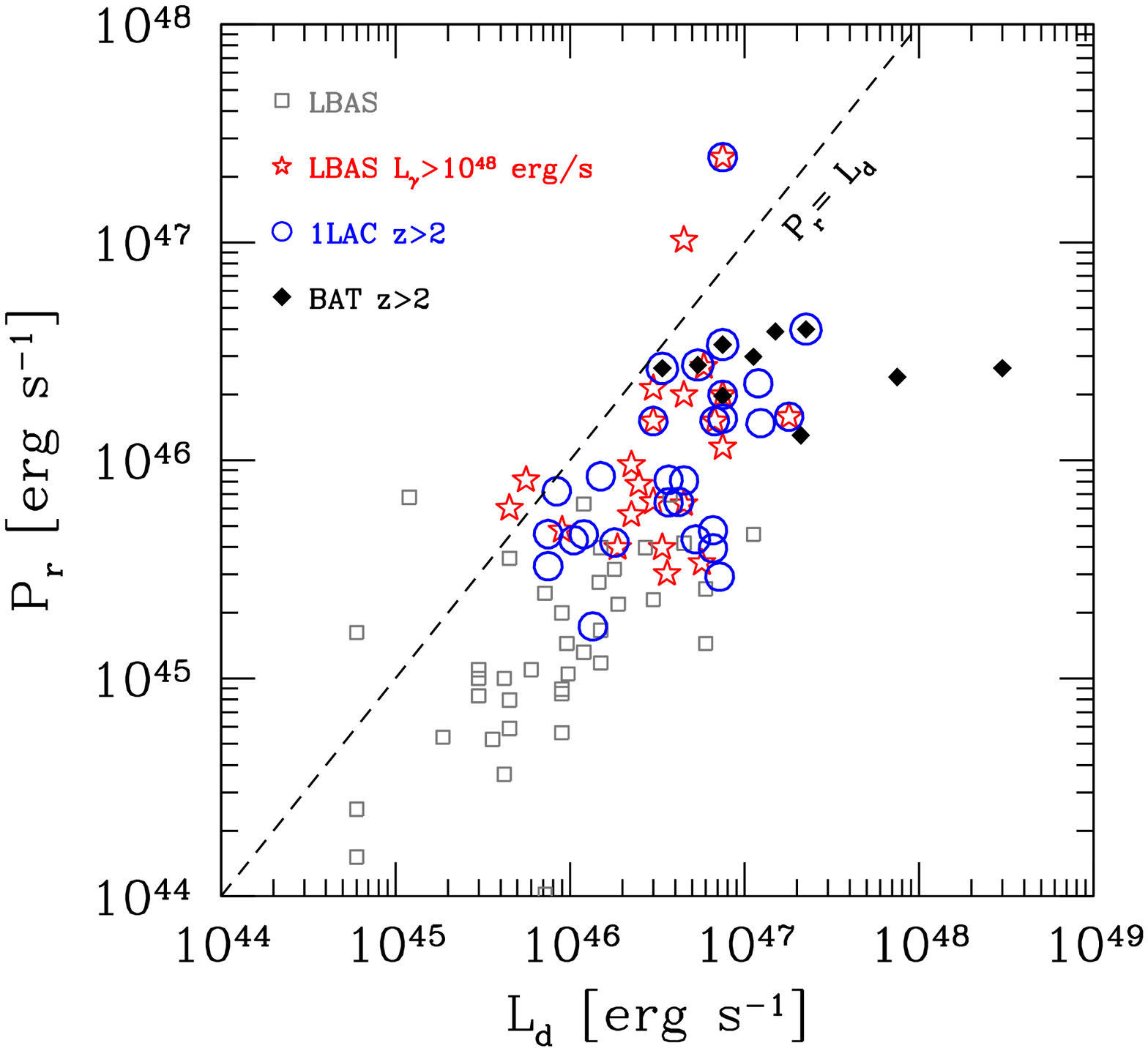,width=8cm,height=7.5cm}
\vskip -0.8 cm
\psfig{figure=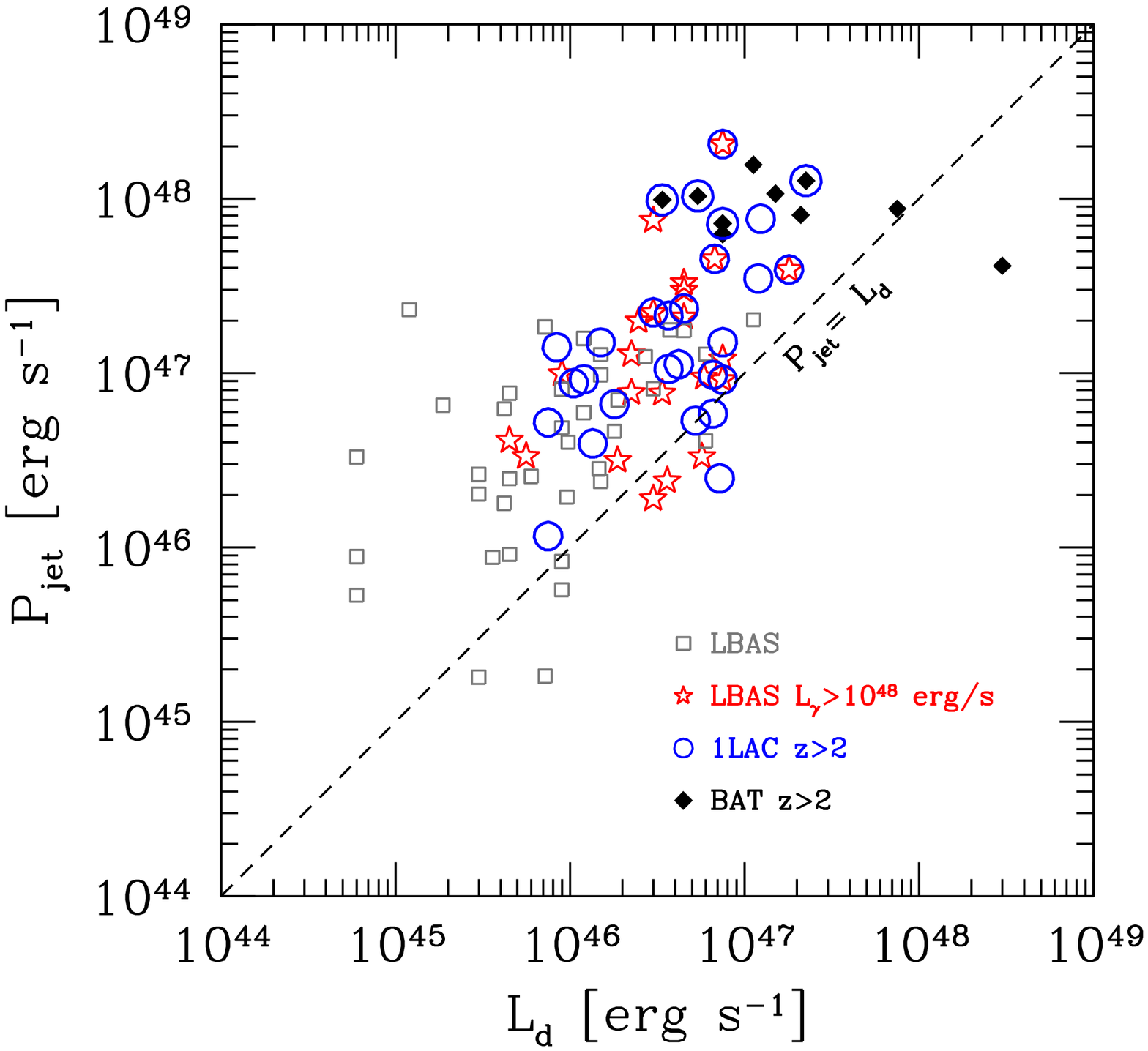,width=8cm,height=7.5cm}
\vskip -0.3 cm
\caption{
Top: the observed $\gamma$--ray luminosity $L_\gamma$
as a function of the accretion
luminosity $L_{\rm d}$ for the LBAS FSRQs (grey squares),
the $z>2$ 1LAC (blue circles)
and {\it Swift}/BAT (diamonds) FSRQs.
The dashed line indicates $L_\gamma=L_{\rm d}$.
Mid: the power $P_{\rm r}$ as a function of $L_{\rm d}$.
$P_{\rm r}$ can be considered as a robust and almost model--independent
lower limit to the jet power.
Bottom: the total jet power $P_{\rm jet}=P_{\rm B}+P_{\rm e}+P_{\rm p}$
as a function of $L_{\rm d}$. 
Almost all sources have $P_{\rm jet}>L_{\rm d}$.
One cold proton per emitting electron is assumed.
}
\label{fpowers}
\end{figure}

\vskip 0.3 cm
\noindent
{\it Disk luminosities ---} We are considering very powerful blazars,
so we do expect large disk luminosities, not only on the basis of 
an expected positive trend between the observed non--thermal (albeit beamed)
and the accretion luminosities, but also on the basis of the observed
luminosities of the broad lines, that should linearly depend on the
accretion power.
What is interesting is that all the FSRQs analyzed up to now (i.e. belonging
to the LBAS sample or to the subset of high redshift 1LAC and BAT samples) 
have a ratio $L_{\rm d}/L_{\rm Edd}$ between 10$^{-2}$ and 1.
This can be seen in the mid panel of Fig. \ref{powerm}, that shows $L_{\rm d}$ as
a function of the derived black hole mass. 
The two dashed lines correspond to the Eddington and 1\% Eddington luminosities.
This confirms the idea of the ``blazars' divide" as a result of the changing
of the accretion mode (Ghisellini, Maraschi \& Tavecchio 2009): 
from the standard Shakura--Sunyaev (appropriate for all FSRQs) to the ADAF--like 
regime (appropriate for BL Lacs).
The $z>2$ blazars analysed here have $L_{\rm d}/L_{\rm Edd}$ ratios
ranging from 0.05 and 0.7.
The exact values of the disk luminosities derived here 
are the frequency integrated bolometric luminosities of
the assumed Shakura--Sunyaev accretion disk model that best
interpolates the data.
On the other hand, any other accretion disk model has to 
fit the data as well, implying that our values of $L_{\rm d}$
are robust, and nearly model--independent, within the limit
of the uncertainties of the observed data.

\vskip 0.3 cm
\noindent
{\it Jet powers ---}  The values listed in Tab. \ref{powers} are very
similar to the values derived for other powerful {\it Fermi} FSRQs.
They are not, however, the absolutely greatest powers found.
This can be seen in the bottom panel of Fig. \ref{powerm},
showing $P_{\rm jet}$ as a function of the black hole mass, and in 
the mid and bottom panels of
Fig. \ref{fpowers}, where we plot the power of the
jet spent in the form of radiation ($P_{\rm r}$) and the total
jet power $P_{\rm jet}$ as a function of $L_{\rm d}$.
We can compare the $z>2$ 1LAC FSRQs with those present in the LBAS
catalogue and the high redshift BAT blazars.
Not surprisingly, we see that in these planes the 1LAC $z>2$ sources follow the
distribution of the most luminous $\gamma$--ray blazars.
Remarkably though, the $z>2$ BAT FSRQs appear to lie at the extreme of the
distributions, being the more powerful in $L_{\rm d}$, and among the most powerful in
$P_{\rm r}$ and $P_{\rm jet}$.
For calculating the power carried by the jet in the form of protons, we re--iterate 
that we have assumed one cold proton per emitting electron:
if there exist a population of cold electrons, and no electron--positron pairs, 
than we underestimate $P_{\rm p}$ and then $P_{\rm jet}$, while 
if there are no cold leptons but there are pairs then we overestimate $P_{\rm p}$.
Finally, protons are assumed cold for simplicity (and ``economy"), but they
could be hot or even relativistic (if, e.g. shocks accelerate not only electrons
but also protons), and in such cases the power is underestimated. 
For a detailed discussion about the presence of electron--positron pairs in
blazars' jets we refer to the discussions in
Sikora \& Madejski (2000), Celotti \& Ghisellini (2008) and G10,
where one can finds arguments limiting the amount of pairs in the jet. 
A few electrons--positrons per proton are possible, but not more.

\vskip 0.3 cm
\noindent
{\it Jet powers vs accretion luminosities ---} Fig. \ref{fpowers} 
shows that the correlations found in G10 between $P_{\rm r}$ and/or 
$P_{\rm jet}$ and $L_{\rm d}$ are confirmed.
We remind the reader that $P_{\rm r}$ and $L_{\rm d}$ are independent
quantities even if the main radiation mechanism is the inverse Compton
process using broad line photons as seeds, that in turn are proportional
to the accretion disk luminosity.
This is because the radiative cooling of the emitting electrons is complete,
implying that the produced jet luminosity becomes independent on the amount
of radiation energy density. 
In other words: in the fast cooling regime
the jet always emits all the energy of its relativistic electrons,
no matter the amount of the luminosity of the accretion disk.

A least square fit returns a chance probability $P=4\times 10^{-8}$ that
$\log P_{\rm r}$ and $\log P_{\rm jet}$ are correlated 
with $\log L_{\rm d}$ (and the correlation are consistent with being linear).
They remain significant also when considering the common redshift dependence,
although the chance probability increases to $P=4\times 10^{-4}$ 
(for the $P_{\rm r}$--$L_{\rm d}$
correlation) and to $P=10^{-3}$ ($P_{\rm jet}$--$L_{\rm d}$).

As expected, the 1LAC blazars at high redshifts are among the most powerful, 
even if there are blazars at lower redshifts with comparable powers.
This can be seen comparing the empty circles, corresponding to the 1LAC blazars of 
our sample, with the LBAS FSRQs of $L_\gamma>10^{48}$ erg s$^{-1}$ (stars) 
and the BAT FSRQs at $z>2$ (black diamonds).
There are a few sources with $P_{\rm r} > L_{\rm d}$, and several with 
$P_{\rm r}\sim L_{\rm d}$. 
The jet in these blazars, only to produce the radiation we see, requires a power 
comparable to (or even larger than) the disk luminosity.
The $P_{\rm r}$ power should be considered a very robust estimate of the 
{\it minimum} jet power: it is robust because it is almost model--independent 
($P_{\rm r} \sim L_\gamma/\Gamma^2$, see G10),
and it is a lower limit because it corresponds to the entire jet power being converted
into radiation at the $\gamma$--ray emitting zone.
Indeed, if there is one proton per emitting electron, the total jet power, dominated
by the bulk motion of cold protons, becomes a factor $\sim 10$ larger than $L_{\rm d}$
(bottom panel of Fig. \ref{fpowers}), with the FSRQs of our sample distributed
in a large portion of the $P_{\rm jet}$--$L_{\rm d}$ plane.

We believe that the relation between both $P_{\rm r}$ and $P_{\rm jet}$ with $L_{\rm d}$ 
is a key ingredient to understand the birth of jets: accretion {\it must} play a key role.

\begin{figure}
\vskip -0.5 cm 
\psfig{figure=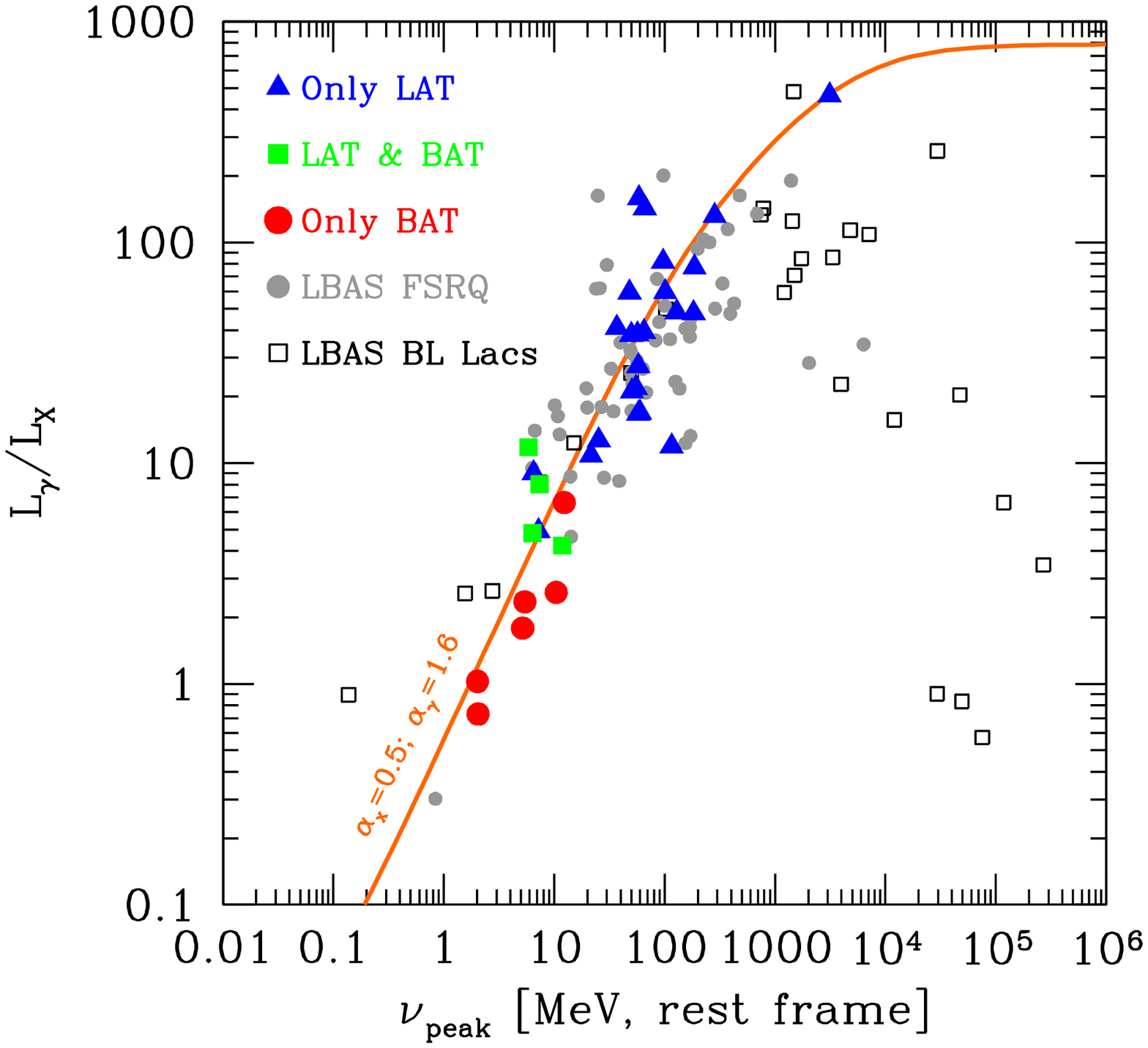,width=9.cm,height=8.cm}
\vskip -0.7 cm
\psfig{figure=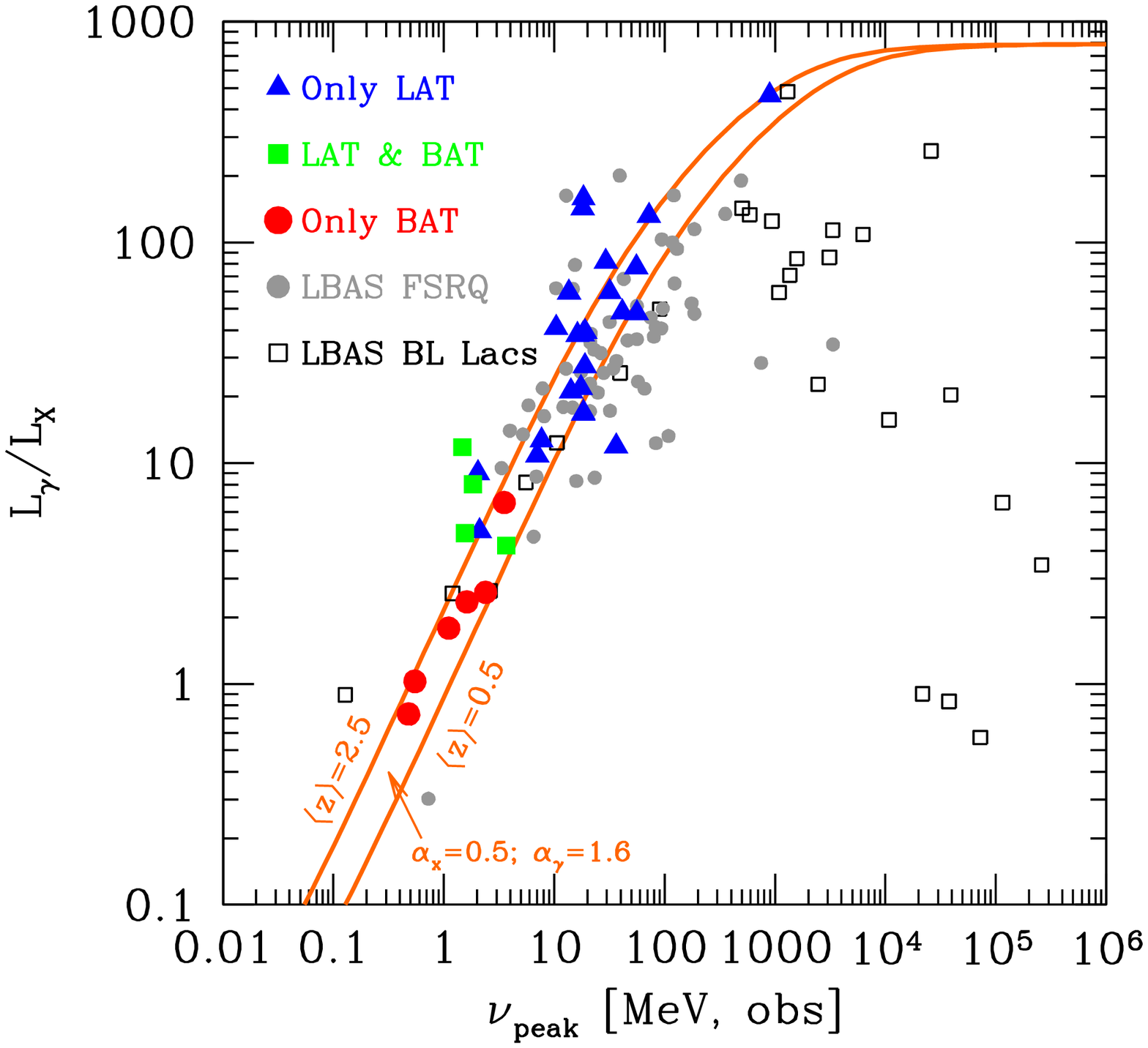,width=9.cm,height=8.cm}
\vskip -0.5 cm
\caption{
Top: The ratio of the (rest frame) luminosities in the 0.1--100 GeV and
15--55 keV bands as a function of the rest frame peak frequency of the high energy bump.
For sources not detected in one of the two bands, we have used the 
corresponding flux resulting from the modelling of their overall SED.
Different symbols are from the sources detected only by {\it Fermi}/LAT,
by both {\it Swift}/BAT and {\it Fermi}/LAT and only by {\it Swift}/BAT, as labelled.
For comparison we also show all FSRQs (grey dots) and BL Lacs (squares) in the
LBAS sample.
The continuous line shows the estimate using the smoothly broken power law 
function (see text) with $\alpha_x=0.5$ and $\alpha_\gamma=1.6$.
Bottom panel: the same, but plotting observed peak frequencies.
The continuous lines are for $\langle z\rangle=2.5$ and $\langle z\rangle=0.5$.
See how the simple function in Eq. \ref{fittino} interpolates well the data of FSRQs.
The X and $\gamma$--ray SED of BL Lacs, instead, is {\it not} due to a single radiation
process, since the hard X--rays are often due to the tail of the synchrotron emission. 
As a consequence, they show the opposite behavior of FSRQs: a smaller $L_X/L_\gamma$
ratio when increasing $\nu_{\rm peak}$, indicating an increasingly 
stronger contribution of the synchrotron flux to the hard X--ray band.
}
\label{peak}
\end{figure}

\begin{figure}
\vskip -0.5 cm 
\psfig{figure=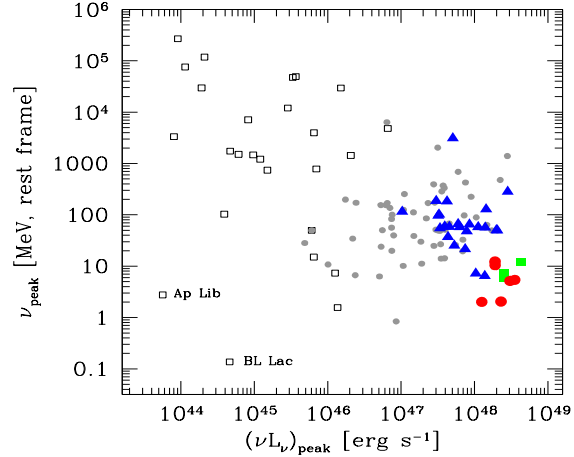,width=8.cm,height=7.cm}
\vskip -0.5 cm
\caption{
The rest frame peak frequency $\nu_{\rm peak}$ of the high energy 
emission as a function of the $\nu_{\rm peak} L_{\nu_{\rm peak}}$ luminosity. 
Symbols as in Fig. \ref{peak}.
A clear trend can be seen when considering BL Lacs and FSRQs together,
while FSRQs only are characterized by a very large dispersion.
High redshifts FSRQs occupy the region of the largest luminosities
and smallest $\nu_{\rm peak}$.
We indicate the two ``outliers": BL Lac and Ap Lib.
}
\label{peak2}
\end{figure}

\vskip 0.3 cm
\noindent
{\it Comparison with other models ---} 
Several groups (Larionov et al. 2008; Marscher et al. 2008, 2010; 
Sikora, Moderski \& Madejski 2008) 
proposed that the emitting region, especially during flares,
is produced at distances from the central black hole of the
order of 10--20 pc (much larger than what we assume) at the expected 
location of a reconfinement shock (e.g. Sokolov, Marscher \& McHardy 2004). 
On the basis of an observed peculiar behaviour of the polarization 
angle in the optical, Marscher et al. (2008) thus suggested that
blobs ejected from the central region are forced by the magnetic field to 
follow a helical path, accounting for the observed rotation of the 
polarization angle in the optical.
Flares (at all wavelengths) correspond to the passage of these blobs
through a standing conical shock, triggered by the compression 
of the plasma in the shock. 
This has important consequences for the variability of the emission: 
since the emission region is located at large
distances from the central engine, its size is large, and the expected
variability timescale cannot be very short.
Assuming $R_{\rm diss} = 15$ pc, a jet aperture angle of $\theta_{\rm jet}=3^\circ$
and $\delta=20$, we find a minimum variability time scale of 
$t_{\rm var}= \theta_{\rm jet} R_{\rm diss} (1+z) /(c \delta)$ of the order of 1.5 (1+z) months,
that for sources at $z>2$ implies a minimum variability timescales of 5--6 months.
The main high energy emission mechanism is still the inverse Compton process,
using as seeds the IR radiation of a surrounding torus (Sikora, Moderski \& Madejski 2008)
with a possible important contribution from jet synchrotron radiation 
(Marscher et al. 2008, 2010).
The main difficulty of these models concerns the expected variability,
predicted to occur on a very long time scales,
if the size of the emitting region is proportional (through e.g. the opening angle of the jet),
to the distance of the source to the black hole.
Instead the observed $\gamma$--ray flux in all strong $\gamma$--ray sources 
(the ones for which a reliable variability behaviour can be established)
varies on much shorter time scales, and factor 2 flux changes can occur even on 3--6 hours 
(see Tavecchio et al. 2010 for 3C 454.3 and PKS 1510--089; 
Bonnoli et al. 2010;
Foschini et al. 2010 and Ackermann et al. 2010 for 3C 454.3; 
Abdo et al. 2009b for PKS 1454--354; 
Abdo et al. 2010c for PKS 1502+105).
This indicates that the source is compact.
This in turn suggests (although it does not prove)
that its location cannot be too far from the black hole.
It then also suggests that it is within the broad line region.
In turn, this suggests that the broad lines are the main seeds for the inverse Compton
scattering process.
Occasionally, though, dissipation could occur further out, where the main
seeds are the infrared photons produced by a putative torus surrounding the
accretion disks.
Since the seed photons have smaller frequencies, in these cases the 
produced high energy spectrum suffers less from possible effects of the decreasing
(with seed frequencies) scattering Klein--Nishina cross section and less from
possible photon--photon interactions leading to electron--positron  pair production.
The decreased importance of both effects may account for high energy spectra
extending, unbroken, up to hundreds of GeV.
These cases should be characterized by a longer variability timescale.

In the model of Marscher et al. (2008) a very short variability timescale
still indicates a very compact emitting region, but nevertheless located at a large
distance from the central black hole.

\subsection{Comparing BAT and LAT high redshift blazars}

Both the 1LAC sample and the BAT 3--years survey have a rather uniform
sky coverage, and both approximate a flux limited sample.
The 1LAC sample has a limiting flux sensitivity that depends on the
$\gamma$--ray spectral index of the sources, but since we are dealing with FSRQs only
(whose $\alpha_\gamma$ is contained in a relatively narrow range), we can consider
this sample as flux limited.
It is then interesting to compare the high redshift blazars 
(all of them are FSRQs) contained in the two samples.
We remind that for blazars with $z>2$, we have 10 FSRQs in the BAT sample,
28 in the 1LAC, and 4 in both.
The X--ray to $\gamma$--ray SED of these sources is very similar:
even if we have only 4 blazars in common, for all sources the
SED has a high energy peak in the $\sim$MeV--100 MeV band,
with $\alpha_x<1$ and $\alpha_\gamma>1$.

Thus the spectrum can be approximated by a broken power law. 
For illustration, consider the smoothly broken power law of the form
\begin{equation}
L(\nu) \propto { (\nu/\nu_{\rm b})^{-\alpha_x} \over
1+(\nu/\nu_{\rm b})^{\alpha_\gamma-\alpha_x} }
\label{fittino}
\end{equation}
If the energy indices $\alpha_x<1$ and $\alpha_\gamma>1$, the peak is at 
$\nu_{\rm peak}=\nu_{\rm b} [(1-\alpha_x)/(\alpha_\gamma-1)]^{1/(\alpha_\gamma-\alpha_x)}$.
With this function we can easily calculate the ratio of the BAT [15--55 keV]
to LAT [0.1--100 GeV] luminosities as a function of $\nu_{\rm peak}$, 
and see if it compares well to the data.
We alert the reader that by ``data" we mean real observed data when the source
has been detected in the X--ray and $\gamma$--ray band, while, when the
detection is missing, we mean the ``data" coming from our fitting model
described in \S 4 (and in more detailed in Ghisellini \& Tavecchio 2009).

The ratio $L_X/L_\gamma$ as a function of $\nu_{\rm peak}$ calculated
using Eq. \ref{fittino} setting $\alpha_x=0.5$ and $\alpha_\gamma=1.6$
is plotted in Fig. \ref{peak} as a grey (orange in the electronic version) line,
together with the points corresponding to high redshift blazars, studied in 
this paper and in G10.
Furthermore, Fig. \ref{peak} reports also the data of all the LBAS  blazars studied
in Ghisellini et al. (2010b). 
These are FSRQs (filled grey circles) and BL Lacs (empty grey squares).
In the top panel $\nu_{\rm peak}$ is in the rest frame of the source, while
in the bottom panel $\nu_{\rm peak}$ is the observed one.
 
Fig. \ref{peak} shows a remarkable agreement between the FSRQ data and the simple
broken power law of Eq. \ref{fittino}, both for high redshifts
and for less distant FSRQs.
BL Lac objects, instead, are not well represented by Eq. \ref{fittino}.
In fact, in many BL Lacs the hard X--rays correspond to the (steep) tail of 
the synchrotron component.
Therefore the X and the $\gamma$--rays are produced by two different mechanisms,
and Eq. \ref{fittino} does not represent their overall high energy SED.
In BL Lacs the importance of X--rays {\it increases} increasing $\nu_{\rm peak}$,
because of the increasing importance of synchrotron flux in the hard X--rays.

Coming back to high redshift FSRQs,
only 4 sources have detection in both bands, while
24 FSRQs are detected only by the LAT and 6 only by the BAT instrument.
Our model explains the large fraction of sources that are detected only in one
of the two instruments as due to the different $\nu_{\rm peak}$ of the sources:
if $\nu_{\rm peak}$ is large (above 10 MeV), $F_\gamma/F_X$ is large and the
source is relatively weak in the BAT band, while if $\nu_{\rm peak}<$10 MeV the source
becomes relatively weak in the LAT band and strong in the BAT.

It is interesting to see if the derived $\nu_{\rm peak}$ correlates with
the bolometric luminosity.
Fig. \ref{peak2} shows that a trend indeed exists: 
more powerful sources have smaller $\nu_{\rm peak}$ 
(we have used $\nu_{\rm peak} L_{\nu_{\rm peak}}$
as a proxy for the bolometric luminosity).
But comparing with all the LBAS FSRQs (grey dots)
we see that the $z>2$ luminous FSRQs belong to a broader (i.e. more scattered)
distribution, and that the high--$z$ BAT blazars are really the most extreme. 

We can conclude that 
i) there is a a trend between the high energy peak and the peak luminosity, 
ii) that this correlation has a large scatter, even if 
iii) the $z>2$ FSRQs show the same trend with less scatter 
(but this may be due to the still small number) and finally 
iv) the $z>2$ blazars in the 3--years BAT survey all lie in the 
highest luminosity, smallest $\nu_{\rm peak}$ part of the plane.

When more BAT detections of high redshift LAT blazars will become available
(and, conversely, when LAT will detect more BAT high--$z$ blazars)
this trend can be tested directly (i.e. without modelling the SED).
The importance of this is two--fold:
first we can estimate in a reasonable way the peak energy of the high energy emission
having the hard X--ray and the $\gamma$--ray luminosities, and second (and more important)
we could conclude that the most powerful blazars can be more easily picked up
trough hard X--ray surveys, as the one foreseen with the {\it EXIST} 
mission (Grindlay et al. 2010).

\section{Summary and discussion}

The total number of $z>2$ blazars with high energy information is 34 
(28 with a LAT detection, 6 with BAT, and 4 with both).
It is still a limited number, the tip of the iceberg of a much larger (and 
fainter) population, but it is derived from two well defined samples (LAT and BAT),
that we can consider as flux limited and coming from two all sky surveys 
(excluding the galactic plane). 
The main result of studying them is that all the earlier findings
concerning the physical parameters of the jet emitting zone,
the jet power, and the correlation between the jet power and the
disk luminosity are confirmed.

They are in agreement with the blazar sequence, i.e. their non--thermal SED
are ``redder" than less luminous blazar, with a large dominance
of their high energy emission over the synchrotron one.
This implies that the disk emission is left unhidden by the synchrotron
flux, and this allows an estimate of the black hole mass and the accretion rate.
The uncertainties associated with these estimates are relatively small 
within the assumption that the thermal component is produced by a 
standard Shakura--Sunyaev disk with an associated non--spinning hole. 
In G10 we argued that in any case the masses are not largely affected
by this assumption, and in particular that in the Kerr case the derived 
masses are not smaller, despite the greater accretion efficiency.
The possibility of an intrinsic collimation of the
disk radiation appears more serious.
If the disk is not geometrically thin, but e.g. a flared disk, then
we expect a disk emission pattern concentrated along the normal
to the disk, i.e. along the jet axis.
We argued previously (Ghisellini et al. 2009) that this can be the 
case of S5 0014+813 at $z=3.366$, an extraordinary luminous
blazars (detected by BAT) with an estimated ``outrageous" black hole
mass of $4\times 10^{10}M_\odot$. 
And indeed we found it to be an ``outlier" in the $P_{\rm r}$--$L_{\rm d}$ plane.
Reverting the argument, it implies that the other FSQRs, obeying a
well defined $P_{\rm r}$--$L_{\rm d}$ trend, should have disk luminosities
with a quasi--isotropic pattern, i.e. standard, not flaring, disks.
The other severe uncertainty on the mass estimation for objects at large 
redshifts is the amount of attenuation of their optical--UV flux, due 
to intervening Lyman--$\alpha$ clouds. 
We have corrected for this assuming an average distribution of clouds, and 
when the attenuation is due to a few thick clouds the expected variance is large.
While we plan to refine such estimates (Haardt et al. in preparation),
it is unlikely that this implies a systematic error on the mass estimates, 
leading to larger values: statistically, the mass distribution should 
not be seriously affected by this uncertainty.

Although affected by the same issues outlined above, also the correlation
between the jet power and the disk luminosity should
resist when a more refined treatment of the Lyman--$\alpha$ absorption
will be available, by the same arguments. 
Therefore we can conclude that the accretion rate is really a fundamental
player in powering the relativistic jet, and we refer the reader to 
Ghisellini et al. (2010b) for a detailed discussion about this finding.
If true, these ideas lead us to suggest that powerful, high redshifts
``true" (i.e. really lineless) BL Lacs do not exist.

Another, at first sight surprising, result of our study is that the
correlation between the peak frequency of the high energy emission and
the $\gamma$--ray luminosity at this peak frequency exists also for 
the $z>2$, highly luminous FSRQs.
It is surprising because in very powerful sources the radiative
cooling is complete (i.e. all electrons with $\gamma$ larger than
an few cool in one light crossing time).
Therefore the electrons responsible for the peak have energies
that are not fixed by radiative cooling, but by the injection
function [i.e. by the $Q(\gamma)$ function given in Eq. 1, and more precisely
by the value of $\gamma_{\rm b}$].
On the other hand, when considering all blazars in the LBAS
sample, we see that the scatter, for large luminosities, is much larger, 
so the apparent correlation for the $z>2$ blazars can be due to small 
statistics.
Clearly, it is a point to investigate further.

\begin{figure}
\vskip -0.5 cm 
\psfig{figure=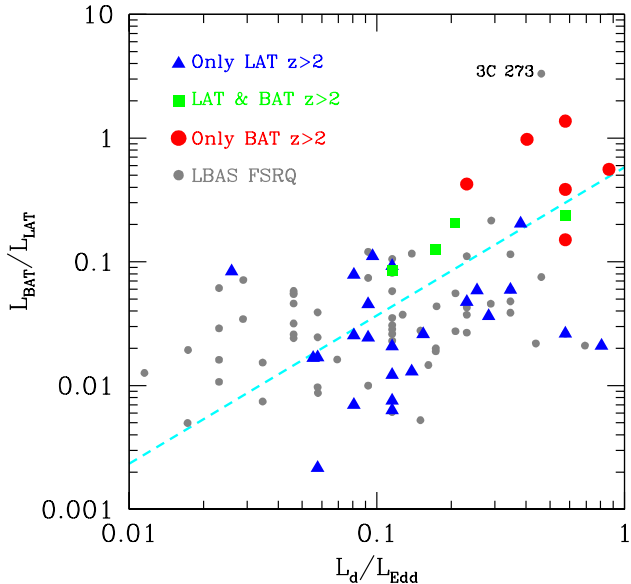,width=9.cm,height=9cm}
\vskip -0.5 cm
\caption{
The ratio between the luminosities in the BAT and LAT energy ranges
(i.e. 15--55 keV and 0.1--10 GeV) as a function of the disk luminosity,
in Eddington units. Big dots are the 6 BAT blazars in the $z>2$ sample 
not detected (yet) by {\it Fermi}/LAT, squares are BAT blazars of
the same sample already present in the 1LAC sample, triangles
are the sources discussed in this paper, and the small grey dots are
all the FSRQs present in the LBAS sample discussed in G10.
The dashed line is the best least square fit (chance probability 
$P=2\times 10^{-6}$).
}
\label{lxlgdotm}
\end{figure}

It is interesting to ask what will be the best strategy to find
the most luminous FSRQs at the largest redshifts.
The interest lies in the link between jet power and disk luminosity:
finding the most powerful jets implies to find the most accreting systems,
hence the heaviest black holes.
Since for each source pointing at Earth (i.e. a blazar) there must
exist $\sim 2\Gamma^2$ similar sources pointing in other directions,
the finding of even a few blazars at large redshifts with a large black 
hole mass can put very interesting constraints on the black hole mass 
function of jetted sources.
This issue has been discussed by us previously (G10),
and we suggested that the existence of the blazar sequence, plus an important
K--correction effect, makes the hard X--ray range the best band where to
search for the high--$z$ heaviest black holes.
Here we re--iterate this suggestion offering a supplementary information.
For each analyzed FSRQs (belonging to the LBAS sample, or in this paper),
we have calculated the ratio between the expected BAT luminosity 
[15--55 keV, rest frame] and the observed [0.1--100 GeV, rest frame] 
LAT luminosity.
This ratio is not an observed quantity, since very few FSRQs (in comparison 
with LAT) have been detected by BAT, but it is a result of the model.
Then, with the same model, we have calculated the disk luminosity in units
of Eddington.
Fig. \ref{lxlgdotm} shows the $L_{\rm BAT}/L_{\rm LAT}$ ratio as a function
of $L_{\rm d}/L_{\rm Edd}$.
We have marked with different symbols the FSRQs analyzed in this paper and
the ones analysed previously (G10).
We have excluded BL Lacs for which we have only an upper limit on the
disk luminosity.
Fig. \ref{lxlgdotm} suggests a clear trend (albeit with some scatter): FSRQs accreting 
close to Eddington emit relatively more in the hard X--rays than above 100 MeV.
Formally, a least square fit returns $(L_{\rm BAT}/L_{\rm BAT}) 
\propto (L_{\rm d}/L_{\rm Edd})^{1.2}$ with a chance probability
$P=2.\times 10^{-6}$ (for 95 objects).
This implies that hard X--ray surveys benefit of a positive bias
when looking for blazars with black holes accreting close to Eddington. 
In turn, at high--$z$, this implies the finding
of the heaviest black holes, since it is very likely that at those
early epochs (e.g. $z>2$) all black holes are accreting close to the
Eddington rate.

\section{Conclusions}

We summarize here our main conclusions:

\begin{itemize}

\item The blazars detected by {\it Fermi} at $z>2$ are all FSRQs, 
with typical ``red" SEDs.

\item These FSRQs are very luminous and powerful, but they are not
at the very extreme of the distribution of luminosity and jet power.

\item These sources have heavy black holes ($M\sim 10^9 M_\odot$)
and accretion luminosities greater than $\sim$10\% Eddington.
When including all FSRQs in the LBAS sample, irrespective of redshift,
the accretion disk luminosities is greater than 1\% Eddington.

\item The trend of redder SED when more luminous (i.e. one of the 
defining characteristics of the blazar sequence) is confirmed,
and it is even present within the relatively small range
of observed luminosity of the $z>2$ blazars.

\item The correlation between the jet power and the disk luminosity is
confirmed and points to a crucial role played by accretion in 
powering the jet.

\item FSRQs with accretion disks closer to the Eddington luminosity
have jets emitting a ``redder" SED, and therefore can be more efficiently
picked up by hard X--ray surveys (such as the one foreseen by {\it EXIST}),
rather than by surveys in the hard $\gamma$--ray band.

\end{itemize}

\section*{Acknowledgments}
This work was partly financially supported by an ASI I/088/06/0) grants.
This research made use of the NASA/IPAC Extragalactic Database (NED) 
which is operated by the Jet Propulsion Laboratory, Caltech, under contract 
with NASA, and of the {\it Swift} public data
made available by the HEASARC archive system.
We also thank the {\it Swift} team for quickly 
approving and performing the requested ToO observations.

\end{document}